\documentclass[onecolumn,aps,prc,showpacs,superscriptaddress]{revtex4}
\usepackage{graphicx}
\usepackage{dcolumn}
\usepackage{bm}

\begin{document}
\title{Electromagnetic interaction in chiral quantum hadrodynamics and
decay of vector and axial-vector mesons }
\author{A.Yu.\ Korchin}
\email{Korchin@kvi.nl}
\affiliation{Laboratory of Theoretical Physics,
University of Gent, B-9000 Gent, Belgium}
\affiliation{NSC `Kharkov Institute of Physics and Technology',
61108 Kharkov, Ukraine}
\author{D.\ Van\ Neck}
\email{Dimitri.VanNeck@rug.ac.be}
\affiliation{Laboratory of Theoretical Physics,
University of Gent, B-9000 Gent, Belgium}
\author{M.\ Waroquier}
\affiliation{Laboratory of Theoretical Physics,
University of Gent, B-9000 Gent, Belgium}

\begin{abstract}
The chiral invariant QHD-III model of Serot and Walecka is applied in the
calculation of some meson properties. The electromagnetic interaction is
included by extending the symmetry of the model to the local $U(1)\times
SU(2)_{R}\times SU(2)_{L}$ group. The minimal and nonminimal contributions
to the electromagnetic Lagrangian are obtained in a new representation of
QHD-III. Strong decays of the axial-vector meson,\ $a_{1}\rightarrow \pi
\rho ,\,\ a_{1}\rightarrow \pi \sigma $, and the electromagnetic decays $
\rho \rightarrow \pi \pi \gamma ,\,\ a_{1}\rightarrow \pi \gamma $ and $\rho
\rightarrow \pi \gamma $ are calculated. The low-energy parameters for the $
\pi-\pi$ scattering are calculated in the tree-level approximation. The
effect of the auxiliary Higgs bosons, introduced in QHD-III in order to
generate masses of the vector and axial-vector mesons via the Higgs
mechanism, is studied as well. This is done on the tree level for $\pi-\pi$
scattering and on the level of one-loop diagrams for the $a_{1}\rightarrow
\pi \gamma $ decay. It is demonstrated that the model successfully describes
some features of meson phenomenology in the non-strange sector.
\end{abstract}

\maketitle

%\keywords{}
%\homepage{}
%\date{\today}

\section{\label{sec:Introduction}Introduction}

Relativistic models built with hadronic degrees of freedom have been very
successful in describing different properties of nuclei and hadrons at low
and intermediate energies (for comprehensive reviews see refs.~\cite
{Ser86,Mei88,Ser97}). In some of these models,
the hadronic Lagrangian has symmetries which are
inspired by the underlying QCD theory.
This allows one to have fewer parameters, thereby
reducing ambiguities in the hadronic models. One of the first models which
incorporated the $SU(2)_{R}\times SU(2)_{L}$ symmetry was the gauged linear
$\sigma $ model (GLSM) developed in ref.~\cite{Lee68}. This model was an
extension of the linear $\sigma $ model and included, in addition to the
pion and scalar meson, the vector $\rho $ and axial-vector $a_{1}$\ mesons
as gauge bosons of the local $SU(2)_{R}\times SU(2)_{L}$ symmetry. The local
symmetry was explicitly broken by the vector-meson mass terms, and
spontaneous symmetry breaking (SSB) by the scalar field led to the mass
splitting between the $\rho $ and $a_{1}.$ This model was elaborated in \cite
{Gas69}, where the current-field identities were established. Later the
model was applied \cite{Ko94} in a description of the meson properties.
Because of some difficulties
additional terms are often included. These terms break further the local
symmetry and introduce additional parameters, which allow for a better
description of the meson observables \cite{Gas69,Ko94,Urb01}.

The QHD-II model, which respects the local $SU(2)_{V}$ isospin symmetry, was
developed in refs.~\cite{Ser79,Ser86}. It was extended in ref.~\cite{Ser92}
by adding the chiral $SU(2)_{A}$ symmetry. This model, called QHD-III, is a
chiral invariant theory based on the local symmetry $SU(2)_{R}\times
SU(2)_{L}$. The $\rho $ and $a_{1}$ mesons are included as the gauge bosons
which are initially massless. The masses are generated through SSB and the
Higgs mechanism. The Lagrangian of QHD-III includes the Lagrangian of GLSM
and the Lagrangian of the Higgs fields. The need for the latter sector of
the model was clearly explained in ref.~\cite{Pre99}: the Higgs mechanism in
GLSM with local symmetry leads to the disappearance of the pion which plays
the role of a would-be Goldstone boson giving its degree of freedom to the
massive $a_{1}$ meson. Therefore the Higgs sector serves to generate the
masses of the $a_1$ and $\rho$ mesons and to preserve the pion as a physical degree of
freedom. Due to the local gauge symmetry the model is renormalizable and
does not require the introduction of cut-off parameters. It is also parity
conserving by construction.

A subtle aspect of QHD-III (but also of GLSM and other hadronic models
including the axial-vector meson) is the presence of a bilinear term mixing
the $a_{1}$ and pion fields. This considerably complicates the
interpretation of the physical particles in the theory as well as
calculations with this Lagrangian. One way to get rid of the mixing was
considered in \cite{Lee68} and later used in other papers \cite
{Gas69,Mei88,Ser92,Ko94}. It consists of a re-definition of the $a_{1}$
field and subsequent wave-function renormalization of the pion field.
The final Lagrangian takes a complicated form with strongly
momentum-dependent vertices. This has undesirable implications in low-energy
meson phenomenology. As examples, we mention the vanishing of the $a_{1}\pi
\rho $ and $\sigma \pi \pi $ vertices at some values of the invariant masses
of the $a_{1}$ and $\sigma$, and the difficulties with the $\rho \pi \pi $
vertex \cite{Urb01}. The authors of \cite{Urb01}, instead of re-defining the
fields, preferred to sum the self-energy generated by the \ $a_{1}-\pi $
mixing to all orders.

An alternative method in the framework of QHD-III was recently suggested in
\cite{Pre99}. This method exploits the freedom in choosing the gauge due to
the local gauge invariance. Originally \cite{Ser92} the so-called unitary
gauge was chosen from the beginning, and the two massless isovector
Goldstone bosons (we will call them $\mathbf{H}$ and $\mathbf{Z}$) were
''gauged away''. This choice leads to the above-mentioned complication with
the mixing. Note that the pseudoscalar boson $\mathbf{Z}$ has the same
quantum numbers $I(J^{P})=1(0^{-})$ as the pion
$\mbox{\boldmath
$\pi$\unboldmath}$ originating from GLSM sector. Therefore the physical pion
field can be chosen as a linear combination of $\mathbf{Z}$ and
$\mbox{\boldmath
$\pi$\unboldmath}$, while the orthogonal combination is decoupled in the
unitary gauge~\cite{Pre99}. The Lagrangian then takes a simpler form, without
complicated momentum-dependent vertices.

In the present paper we apply this new representation of QHD-III in the
calculation of some meson properties. First, we include the electromagnetic
(EM) interaction in this model. This is done via an extension of the
symmetry of QHD-III to the local $U(1)\times SU(2)_{R}\times SU(2)_{L}$
group. We use an arbitrary gauge where all eight Higgs fields are initially
present. The EM interaction in both sectors of the model is obtained. After
an appropriate diagonalization of the $\mathbf{Z}$ and
$\mbox{\boldmath$\pi$\unboldmath}$ fields and fixing the gauge along
the lines of ref.~\cite{Pre99}, we obtain the EM interaction in terms
of the physical pion field.
The final Lagrangian includes the minimal EM interaction, as well as the
(nonminimal) EM interaction with the intrinsic magnetic moment of the $\rho$
and $a_1$ mesons.

We then study the strong and EM decays of the vector and axial-vector
mesons. Some of the decays can be calculated on the ``tree-graph" level,
while others require a calculation beyond the tree level. In particular,
we calculate the width of the following decays: $a_{1}\rightarrow \pi \rho
\,,$ $a_{1}\rightarrow \pi \sigma $ and $\rho ^{0}\rightarrow \pi ^{+}\pi
^{-}\gamma .$ We also address the issue of the width of the scalar $\sigma $
meson, in view of the interest \cite{Ko94} in this subject. The matrix
elements for the above decays are given directly by the corresponding
vertices in the Lagrangian. In order to calculate the decays $\
a_{1}^{+}\rightarrow \pi ^{+}\gamma $ and $\ \rho ^{+}\rightarrow \pi
^{+}\gamma $ we need to include loop diagrams. In particular, the
$a_{1}^{+}\rightarrow \pi ^{+}\gamma $ process is described by a large number
of one-loop diagrams, which can be grouped according to the intermediate
state in the diagram.  The
diagrams where only vector or axial vector mesons
in the loop are present are not yet included in this exploratory study.

All matrix elements of the EM processes turn out to be
finite, due to a cancellation of divergencies between different amplitudes.
The fulfillment of EM gauge invariance serves as a check of the calculation.
The decay widths of these processes are listed in the PDG reviews, and
we compare the model predictions with experimental values.

To determine the parameters of the model we fix the strong-coupling constant
$g_{\rho }$ from the $\rho \rightarrow \pi \pi $ decay. All other parameters
are strongly correlated, once the masses of the particles are taken equal to
their experimental values. Only the mass $m_{\sigma }$\ of the $\sigma $
meson and the mass $m_{H}$ of the Higgs particles remain unconstrained. The
mass $m_{H}$ is taken to infinity in the calculation. As an
additional test of the model we calculate the low-energy parameters for $\pi
-\pi $ scattering. Because of some unusual features of the model, such as
the presence of Higgs mesons and a suppression of the $\pi \pi \sigma $
interaction, it is not \textit{a priori} clear whether the model can
reasonably describe the experiment. The scattering lengths and effective
ranges for the $S-$ and $P-$waves are calculated on the tree level and
compared with the data and other approaches.

The paper is organized as follows. In Sect.~\ref{sec:EM} the Lagrangians of
the EM and strong interactions are obtained. We start with the Higgs sector
in Sect.~\ref{subsec:EM-Higgs}. The GLSM is briefly discussed in Sect.~\ref
{subsec:Sigma}. In Sect.~\ref{subsec:Removing} the procedure for removing
the mixing terms in the Lagrangian is described. The final EM and
strong-interaction Lagrangians in terms of the physical pion field are presented in
Sect.~\ref{subsec:EM-physical}. In Sect.~\ref{subsec:Strong decays} the
widths of the strong decay of the mesons are calculated, and in Sect.~\ref
{subsec:EM decays} we consider the EM decay of the mesons. Results are
compared with experiment. Pion-pion scattering at low energies is studied in
Sect.~\ref{sec:pi-pi}. In Sect.~\ref{sec:Discussion} we discuss the results and prospects,
and draw conclusions. In Appendix~\ref{App:A} we outline the derivation of
the Lagrangian, originating from the Higgs
sector. Explicit expressions are given for the Lagrangian in GLSM. Finally,
Appendix~\ref{App:B} contains details of the calculation of the one-loop
integrals.

%%%%%%%%%%%%%%%%%%%%%%%%%%%%%%%%%%%%%%%%%%%%%%%%%%%%%%%
%%%%%%%%%%%%%%%%%%%%%%%%%%%%%%%%%%%%%%%%%%%%%%%%%%%%%%%

\section{\label{sec:EM} Electromagnetic interaction in chiral quantum
hadrodynamics}

\subsection{\label{subsec:EM-Higgs} Lagrangian of electromagnetic
interaction in the Higgs sector}

In this section we discuss the EM interaction in the framework of chiral
quantum hadrodynamics (QHD-III). The strong Lagrangian \cite{Ser92} consists
of the GLSM Lagrangian $\mathcal{L}_{N\pi \sigma \omega }$ and the Higgs
part,
\begin{equation}
\mathcal{L}_{QHD-III}=\mathcal{L}_{N\pi \sigma \omega }+\mathcal{L}_{H},
\label{Eq.1}
\end{equation}
where $\mathcal{L}_{N\pi \sigma \omega }$ will be discussed in the next
section, and
\begin{equation}
\mathcal{L}_{H}=(D_{\mu }\Phi _{R})^{\dagger }(D^{\mu }\Phi _{R})+(D_{\mu
}\Phi _{L})^{\dagger }(D^{\mu }\Phi _{L})-V_{H}(\Phi _{R},\Phi _{L}),
\label{Eq.2}
\end{equation}
with the potential
\begin{equation}
V_{H}(\Phi _{R},\Phi _{L})=\frac{\lambda _{H}}{4}[(\Phi _{R}{}^{\dagger
}\Phi _{R})^{2}+(\Phi _{L}{}^{\dagger }\Phi _{L})^{2}]-\mu _{H}^{2}(\Phi
_{R}{}^{\dagger }\Phi _{R}+\Phi _{L}{}^{\dagger }\Phi _{L}).  \label{Eq.3}
\end{equation}

The complex doublets of spinless fields, $\Phi _{R},\Phi _{L},$ transform as
the spinor representation of $\ SU(2)$. The covariant derivatives $D_{\mu
\text{ }}$are expressed in terms of the right and left isovector gauge
fields $\mathbf{r}_{\mu }=(\mbox{\boldmath
$\rho$\unboldmath}_{\mu }^{\prime }+\mathbf{a}_{\mu })/\sqrt{2}$ and
$\mathbf{l}_{\mu }=(\mbox{\boldmath
$\rho$\unboldmath}_{\mu }^{\prime }-\mathbf{a}_{\mu })/\sqrt{2}.$ The
Lagrangian is symmetrical under the local $SU(2)_{R}\times SU(2)_{L}$ gauge
transformations (for more details see \cite{Ser92}).

To include the EM interaction we extend the model by adding the gauge $U(1)$
symmetry of hypercharge. The method is formally equivalent to that in the
theory of Glashow, Weinberg and Salam (GWS) of electroweak interactions
(see, e.g., \cite{Pes95} Ch.20.2, and also \cite{Ser86}). The hypercharge
$Y_{H}=1$ is assigned to the scalar fields, and the covariant derivatives
acting on $\Phi _{R}$ and $\Phi _{L}$\ take the form
\begin{equation}
D_{\mu }\Phi _{R/L}=[\partial _{\mu }+ig^{\prime}_{\rho }\frac{\mbox{\boldmath$\tau$
\unboldmath }}{2}(\mbox{\boldmath $\rho$\unboldmath}_{\mu }^{\prime } \pm
\mathbf{a}_{\mu })+\frac{i}{2}e^{\prime}A_{\mu }^{\prime }]\Phi _{R/L},  \label{Eq.4}
\end{equation}
where $A_{\mu }^{\prime }$ is the EM field,  $g_{\rho }^{\prime}$ and $e^{\prime}$
are the strong and EM coupling constants. The
fields $\mbox{\boldmath$\rho$\unboldmath}_{\mu }^{\prime }\;$and$\;\mathbf{a}
_{\mu }$ are associated with the vector (isovector) meson $\rho (770)$, and
axial-vector (isovector) meson $a_{1}(1260).$ We should also add the free
Lagrangians of all vector fields
\begin{equation}
\mathcal{L}_{\gamma \rho a}^{(0)}=-\frac{1}{4}
\mbox{\boldmath
$\rho$\unboldmath}_{\mu \nu }^{{\prime }2}-\frac{1}{4}\mathbf{a}_{\mu \nu
}^{2}-\frac{1}{4}(\partial _{\mu }A_{\nu }^{\prime }-\partial _{\nu }A_{\mu
}^{\prime })^{2},  \label{Eq.5}
\end{equation}
where
\begin{eqnarray}
\mbox{\boldmath $\rho$\unboldmath}_{\mu \nu }^{\prime } &=&\partial _{\mu }
\mbox{\boldmath $\rho$\unboldmath}_{\nu }^{\prime }-\partial _{\nu }
\mbox{\boldmath $\rho$\unboldmath}_{\mu }^{\prime }-g^{\prime}_{\rho }(
\mbox{\boldmath $\rho$\unboldmath}_{\mu }^{\prime }\times
\mbox{\boldmath
$\rho$\unboldmath}_{\nu }^{\prime })-g^{\prime}_{\rho }(\mathbf{a}_{\mu }\times
\mathbf{a}_{\nu }),  \nonumber \\
\mathbf{a}_{\mu \nu } &=&\partial _{\mu }\mathbf{a}_{\nu }-\partial _{\nu }
\mathbf{a}_{\mu }-g^{\prime}_{\rho }(\mbox{\boldmath $\rho$\unboldmath}_{\mu
}^{\prime }\times \mathbf{a}_{\nu })-g^{\prime}_{\rho }(\mathbf{a}_{\mu }\times
\mbox{\boldmath $\rho$\unboldmath}_{\nu }^{\prime }).  \label{Eq.6}
\end{eqnarray}
In these expressions we included primes on $A_{\mu }$ and $
\mbox{\boldmath
$\rho$\unboldmath}_{\mu }$, anticipating that these are not yet the fields
of the physical photon and $\rho $ meson, but will be redefined.

Due to the chosen form of the potential $V_{H} $  in Eq.(\ref{Eq.3}), the
masses of the vector and axial-vector mesons are generated via the Higgs
mechanism, as suggested in ref.~\cite{Ser92}. The fields $\Phi _{R}$ and $
\Phi _{L}$ acquire a nonzero vacuum expectation value (VEV)
\begin{equation}
\langle \Phi _{R}\rangle =\langle \Phi _{L}\rangle =\frac{1}{2}\left(
\begin{array}{c}
0 \\
u
\end{array}
\right) ,  \label{Eq.7}
\end{equation}
where the value of $u$ will be specified later. We now define the eight
Higgs fields via
\begin{equation}
\Phi _{R/L}=\frac{1}{2}[u+\eta \pm \zeta +i\mbox{\boldmath $\tau$\unboldmath}
(\mathbf{H \pm Z})]\left(
\begin{array}{c}
0 \\
1
\end{array}
\right) .  \label{Eq.8}
\end{equation}
The fields $\eta $ and $\mathbf{H}$ are scalars, whereas $\zeta $ and $
\mathbf{Z}$ are pseudoscalars under the parity transformation,
\begin{eqnarray}
\mathcal{P}\eta (t,\mathbf{x})\mathcal{P}^{-1} &=&\eta (t,-\mathbf{x}),\;\;\
\ \;\ \ \;\ \ \;\mathcal{P}\mathbf{H}(t,\mathbf{x})\mathcal{P}^{-1}=\mathbf{H
}(t,-\mathbf{x}),\;\;\;\;\;  \nonumber \\
\mathcal{P}\zeta (t,\mathbf{x})\mathcal{P}^{-1} &=&-\zeta (t,-\mathbf{x}
),\;\;\ \ \;\ \ \;\;\mathcal{P}\mathbf{Z}(t,\mathbf{x})\mathcal{P}^{-1}=-\mathbf{Z}
(t,-\mathbf{x})\,,  \label{Eq.9}
\end{eqnarray}
so that the fields $\Phi _{R}$ and $\Phi _{L}$ satisfy the relations
\begin{equation}
\mathcal{P}\Phi _{R}(t,\mathbf{x})\mathcal{P}^{-1}=\Phi _{L}(t,-\mathbf{x}
),\;\;\ \ \ \ \;\;\;\mathcal{P}\Phi _{L}(t,\mathbf{x})\mathcal{P}^{-1}=\Phi
_{R}(t,-\mathbf{x})\,.  \label{Eq.10}
\end{equation}
Eq.(\ref{Eq.10}) is the condition that the model is parity conserving \cite
{Ser92}.

The Lagrangian can now be rewritten in terms of the fields $
\mbox{\boldmath
$\rho$\unboldmath}_{\mu }^{\prime },\mathbf{a}_{\mu },A_{\mu }^{\prime
},\eta ,\zeta ,\mathbf{H}$ and $\mathbf{Z}.$ We insert Eq.(\ref{Eq.4}) in
the Lagrangian (\ref{Eq.2}) and use the
representation of Eq.(\ref{Eq.8}). In the derivation there appears a mixing
between $A_{\mu }^{\prime }$ and the 3d component of $
\mbox{\boldmath
$\rho$\unboldmath}_{\mu }^{\prime }$, which requires a re-definition of
these fields. One can introduce new physical fields (without primes),
\begin{equation}
\left(
\begin{array}{c}
{\rho }_{\mu }^{3} \\
{A}_{\mu }
\end{array}
\right) =\left(
\begin{array}{cc}
\cos \theta & -\sin \theta \\
\sin \theta & \cos \theta
\end{array}
\right) \left(
\begin{array}{c}
\rho _{\mu }^{{\prime }3} \\
A_{\mu }^{\prime }
\end{array}
\right) ,  \label{Eq.11}
\end{equation}
with the mixing angle defined through
$\tan \theta =e^{\prime}/g^{\prime}_{\rho },$ and
rewrite the Lagrangian in terms of the physical fields.
The covariant derivatives now read
\begin{equation}
D_{\mu }\Phi _{R/L}=[\partial _{\mu }
 \pm \frac{i}{2} g^{\prime}_{\rho }
\mbox{\boldmath$\tau$\unboldmath} \mathbf{a}_{\mu } +
\frac{i}{2} g^{\prime}_{\rho } ({ \tau^{1}}\rho^{1}_{\mu }
+ { \tau^{2}}\rho^{2}_{\mu } )
+\frac{i}{2} g_{\rho}({\tau^3} - 2 Q_H \sin^{2}{\theta})\rho^{3}_{\mu}
+i e Q_H A_{\mu }]\Phi _{R/L}\,,
\label{Eq.13}
\end{equation}
where we introduced the coupling of the neutral $\rho$ meson, $g_{\rho} \equiv
g^{\prime}_{\rho}/ \cos{\theta}$, and the electric charge of the proton,
$e \equiv g^{\prime}_{\rho} \sin{\theta} = e^{\prime} \cos{\theta}$.
The charge operator is given by $Q_{H}= (1+\tau _{3})/2$; it is seen that it
yields zero when acting on the vacuum. The latter condition is crucial
to ensure that the photon does not acquire mass due to SSB.

For the physical values of the couplings we have $\theta \ll 1$ and,
up to $\mathcal{O}(e/g_{\rho})$, we use the substitutions
\begin{equation}
\rho _{\mu }^{{\prime }3}\rightarrow {\rho }_{\mu }^{3}+\frac{e}{g_{\rho }}{A}
_{\mu },\;\;\;\ \ \;\ \ \;
A_{\mu }^{\prime }\rightarrow {A}_{\mu }-\frac{e}{
g_{\rho }}{\rho }_{\mu }^{3},\;\ \ \ \ \ \ \ \ \ \ \ \rho _{\mu }^{{\prime }
1,2} \rightarrow {\rho }_{\mu }^{1,2},
\label{Eq.12}
\end{equation}
without distinguishing $g^{\prime}_{\rho }$ from $g_{\rho }$,
and $e^{\prime}$ from $e$. Eq.(\ref{Eq.13})
simplifies correspondingly, to this order.

The derivation of the Lagrangian is tedious, and some details are collected
in Appendix~\ref{App:A}. The result can be written as a sum of the EM and
strong-interaction parts,
\begin{equation}
\mathcal{L}_{H}=\mathcal{L}_{H}^{em}+\mathcal{L}_{H}^{str},  \label{Eq.14}
\end{equation}
where the EM\ Lagrangian is
\begin{eqnarray}
\mathcal{L}_{H}^{em}&=&-\frac{1}{4}(\partial _{\mu }A_{\nu }-\partial _{\nu
}A_{\mu })^{2}+\mathcal{L}_{H}^{em,{min}}+\mathcal{L}_{H}^{em,{nonmin}}\, ,
\label{Eq.15} \\
\mathcal{L}_{H}^{em,{min}} &=& -eA_{\mu }J_{H}^{\mu }\, ,  \label{Eq.16}
\end{eqnarray}
\begin{eqnarray}
J_{H}^{\mu } &=&(\mathbf{H}\times \partial ^{\mu }\mathbf{H}+\mathbf{Z}
\times \partial ^{\mu }\mathbf{Z})_{3}+m_{\rho }(\mathbf{Z}\times \mathbf{a}
^{\mu }+\mathbf{H}\times \mbox{\boldmath $\rho$\unboldmath}^{\mu })_{3}
\nonumber \\
&& +\frac{1}{2}g_{\rho }[\mathbf{H}\times (\mathbf{H}\times
\mbox{\boldmath
$\rho$\unboldmath}^{\mu })+\mathbf{Z}\times (\mathbf{Z}\times
\mbox{\boldmath
$\rho$\unboldmath}^{\mu }) +\mathbf{Z}\times (\mathbf{H}\times \mathbf{a}
^{\mu })+\mathbf{H}\times (\mathbf{Z}\times \mathbf{a}^{\mu })  \nonumber \\
&& +\eta (\mathbf{Z}\times \mathbf{a}^{\mu }+\mathbf{H}\times
\mbox{\boldmath $\rho$\unboldmath}^{\mu })+\zeta (\mathbf{H}\times \mathbf{a}
^{\mu }+\mathbf{Z}\times \mbox{\boldmath $\rho$\unboldmath}^{\mu })]_{3} +(
\mbox{\boldmath $\rho$\unboldmath}^{\mu \nu }\times
\mbox{\boldmath
$\rho$\unboldmath}_{\nu }+\mathbf{a}^{\mu \nu }\times \mathbf{a}_{\nu })_{3},
\label{Eq.17}
\end{eqnarray}
\begin{equation}
\mathcal{L}_{H}^{em,{nonmin}} =\frac{e}{2}(\partial _{\mu }A_{\nu }-\partial
_{\nu }A_{\mu })(\mbox{\boldmath $\rho$\unboldmath}^{\mu }\times
\mbox{\boldmath $\rho$\unboldmath}^{\nu }+\mathbf{a}^{\mu }\times \mathbf{a}
^{\nu })_{3}.  \label{Eq.18}
\end{equation}
It is seen
that in the arbitrary gauge there is a contribution originating from the
fields $\ \mathbf{H}$ and $\mathbf{Z}$. If they are omitted from the
beginning then only the last term in Eq.(\ref{Eq.17}) would remain. In fact
the field $\mathbf{Z}$ may survive even in the unitary gauge (see Sect.\ref
{subsec:Removing}) and contribute to the EM current. To clarify this point
we need to consider explicitly the second sector of the model -- GLSM (Sect.~
\ref{subsec:Sigma}). It is also worthwhile to notice the nonminimal EM
interaction in Eq.(\ref{Eq.18}), which comes from the free $\rho $-meson
Lagrangian after making the substitutions of Eq.(\ref{Eq.12}).

The strong-interaction Lagrangian in Eq.(\ref{Eq.14}) has the following structure
\begin{equation}
\mathcal{L}_{H}^{str}=\mathcal{L}_{HZ}^{(0)}+\mathcal{L}_{\eta \zeta }^{(0)}+
\mathcal{L}_{\rho a}^{(0)}+\mathcal{L}_{H}^{int}+m_{\rho }(\mathbf{a}_{\mu
}\partial ^{\mu }\mathbf{Z}+\mbox{\boldmath $\rho $\unboldmath}_{\mu
}\partial ^{\mu }\mathbf{H}),  \label{Eq.19}
\end{equation}
where $\mathcal{L}_{HZ}^{(0)},\,\mathcal{L}_{\eta \zeta }^{(0)}$ and $
\mathcal{L}_{\rho a}^{(0)}$ are the free Lagrangians of the massless
Goldstone bosons$\ \mathbf{H,Z}$, the massive Higgs bosons $\eta ,\zeta $,
and the gauge bosons $\mbox{\boldmath
$\rho$\unboldmath}_{\mu },\mathbf{a}_{\mu }$. We have respectively,
\begin{eqnarray}
\mathcal{L}_{HZ}^{(0)}+\mathcal{L}_{\eta \zeta }^{(0)} &=&\frac{1}{2}
(\partial _{\mu }\mathbf{H})^{2}+\frac{1}{2}(\partial _{\mu }\mathbf{Z})^{2}+
\frac{1}{2}[(\partial _{\mu }\eta )^{2}-m_{H}^{2}\eta ^{2}]+\frac{1}{2}[
(\partial _{\mu }\zeta )^{2}-m_{H}^{2}\zeta ^{2}], \   \label{Eq.20} \\
\mathcal{L}_{\rho a}^{(0)} &=&-\frac{1}{4}\mbox{\boldmath $\rho$\unboldmath}
_{\mu \nu }^{2}+\frac{1}{2}m_{\rho }^{2}\mbox{\boldmath $\rho$\unboldmath}
_{\mu }^{2}-\frac{1}{4}\mathbf{a}_{\mu \nu }^{2}+\frac{1}{2}m_{\rho }^{2}
\mathbf{a}_{\mu }^{2}.  \label{Eq.21}
\end{eqnarray}
The expression for the interaction term $\mathcal{L}_{H}^{int}$ is
complicated and given in Eqs.(\ref{Eq.A-Higgs-strong}) and (\ref{Eq.A-V-Higgs})
of Appendix~\ref{App:A}. The last term in Eq.(\ref{Eq.19}) describes the SSB
induced mixing of $\mbox{\boldmath $\rho $\unboldmath}_{\mu }$ with $\mathbf{
H}$ and of $\mathbf{a}_{\mu }$ with $\mathbf{Z}$. This term will be dealt
with in Sect.~\ref{subsec:Removing}. The mass of the $\rho $ and $a_{1}$
mesons
\footnote{
In order $O(e^{2}/g^{2}_{\rho})$ there would appear mass splitting between the neutral
and charged $\rho $ mesons, $m_{\rho_{\pm}}=m_{\rho_{0}} \cos{\theta}.$},
the mass of the $\eta $ and $\zeta $, the VEV $u$, and the parameters of the
potential are related via
\begin{equation}
m_{\rho }=\frac{1}{2}g_{\rho }u,\;\ \ \ \ \;\ \ \ \ \ \ \ \ \ \ \ \ m_{H}=
\sqrt{2}\mu _{H},\;\ \ \ \ \ \ \ \ \ \ \ \;\ \ \ \ \lambda _{H}=\Big(\frac{
m_{H}g_{\rho }}{m_{\rho }}\Big)^{2}.  \label{Eq.22}
\end{equation}

%%%%%%%%%%%%%%%%%%%%%%%%%%%%%%%%%%%%%%%%%%%%
%%%%%%%%%%%%%%%%%%%%%%%%%%%%%%%%%%%%%%%%%%%%

\subsection{\label{subsec:Sigma} Electromagnetic interaction in the gauged
linear sigma model}

The Lagrangian of GLSM \cite{Lee68,Gas69,Ser92} can be written in terms of
the fields of the nucleon ($N$), pion ($\mbox{\boldmath $\pi$\unboldmath}$),
scalar meson ($\phi $), and vector mesons ($\omega _{\mu },\;
\mbox{\boldmath
$\rho$\unboldmath}_{\mu }^{\prime }$ and $\mathbf{a}_{\mu }$) as follows
\begin{equation}
\mathcal{L}_{N\pi \sigma \omega }=\bar{N}[i\gamma ^{\mu }D_{\mu }-g_{\pi
}(\phi +i\gamma _{5}\mbox{\boldmath $\tau$\unboldmath}
\mbox{\boldmath
$\pi$\unboldmath})]N+\frac{1}{2}[(\Delta _{\mu }
\mbox{\boldmath
$\pi$\unboldmath})^{2}+(\Delta _{\mu }\phi )^{2}]+\frac{1}{2}m_{\omega
}^{2}\omega _{\mu }^{2}-\frac{1}{4}\omega _{\mu \nu }^{2}-V(\phi ,
\mbox{\boldmath
$\pi$\unboldmath})+\mathcal{L}_{SB},  \label{Eq.23}
\end{equation}
where the covariant derivatives acting on the nucleon, pion and scalar
fields are defined respectively as
\begin{eqnarray}
D_{\mu }N &=&[\partial _{\mu }+ig^{\prime}_{\rho }\frac{
\mbox{\boldmath
$\tau$\unboldmath}}{2}(\mbox{\boldmath $\rho$\unboldmath}_{\mu }^{\prime
}+\gamma _{5}\mathbf{a}_{\mu })+ig_{\omega }\omega _{\mu }]N,
 \label{Eq.24}
\\
\Delta _{\mu }\mbox{\boldmath $\pi$\unboldmath} &=&\partial _{\mu }
\mbox{\boldmath $\pi$\unboldmath}+g^{\prime}_{\rho }\mbox{\boldmath $\pi$\unboldmath}
\times \mbox{\boldmath $\rho$\unboldmath}_{\mu }^{\prime }
-g^{\prime}_{\rho }\phi
\mathbf{a}_{\mu },\,\ \ \ \ \ \ \ \ \ \ \ \ \ \ \ \ \ \ \ \ \ \ \ \Delta
_{\mu }\phi =\partial _{\mu }\phi +g^{\prime}_{\rho }\mbox{\boldmath $\pi$\unboldmath}
\mathbf{a}_{\mu },\;\;\ \ \ \ \ \ \ \;\;\ \   \label{Eq.25}
\end{eqnarray}
and the kinetic energy of the \ $\omega $ meson is expressed through the
tensor $\omega _{\mu \nu }=\partial _{\mu }\omega _{\nu }-\partial _{\nu
}\omega _{\mu }.\;$The potential energy term is
\begin{equation}
V(\phi ,\mbox{\boldmath $\pi$\unboldmath})=\frac{1}{4}\lambda (\phi ^{2}+
\mbox{\boldmath $\pi$\unboldmath}^{2})^{2}-\frac{1}{2}\mu ^{2}(\phi ^{2}+
\mbox{\boldmath $\pi$\unboldmath}^{2}).  \label{Eq.26}
\end{equation}
Note that there are no mass terms for the nucleon, $\rho $ and $a_{1}$
mesons, whereas a mass term is present for the isoscalar $\omega$. This
Lagrangian is invariant under local $SU(2)_{R}\times SU(2)_{L}$
transformations, apart from a possible explicit symmetry-breaking term $
\mathcal{L}_{SB}=c\phi $ generating the pion mass.

The EM interaction is included by changing $D_{\mu }N$ $\ $to $\ (D_{\mu }+
\frac{i}{2}e^{\prime}Y_{N}A_{\mu }^{\prime })N$, where the nucleon hypercharge $Y_{N}$
is taken equal to unity. We also have to make the substitutions of Eq.(\ref
{Eq.12}). The covariant derivative for the nucleon, in the order
$\mathcal{O}(e/g_{\rho})$, takes the form
\begin{equation}
D_{\mu }N\rightarrow \lbrack \partial _{\mu }+ig_{\rho }\frac{
\mbox{\boldmath $\tau$\unboldmath}}{2}(\mbox{\boldmath $\rho$\unboldmath}
_{\mu }+\gamma _{5}\mathbf{a}_{\mu })+ig_{\omega }\omega _{\mu }+\frac{i}{2}
e(1+\tau _{3})A_{\mu }]N\,,  \label{Eq.27}
\end{equation}
and the electric charge of the nucleon is $e Q_{N}=e(T_{3}+Y_{N}/2)$ \ in
accordance with the Gell-Mann - Nishijima relation.

The nucleon mass is generated via the SSB, if $\lambda >0$ and $\mu ^{2}>0$
in Eq.(\ref{Eq.26}). The scalar field acquires a nonzero VEV $\langle \phi
\rangle =v,$ and after redefining the sigma field via $\phi =v+\sigma $ we
obtain the following Lagrangian of the EM interaction
\begin{equation}
\mathcal{L}_{N\pi \sigma }^{em}=-eA_{\mu }J_{N\pi \sigma }^{\mu }
\label{Eq.28}
\end{equation}
with the EM current
\begin{equation}
J_{N\pi \sigma }^{\mu }=\bar{N}\gamma ^{\mu }\frac{1}{2}(1+\tau _{3})N+[
\mbox{\boldmath $\pi$\unboldmath}\times \partial ^{\mu }
\mbox{\boldmath
$\pi$\unboldmath}-g_{\rho }(v+\sigma )\mbox{\boldmath $\pi$\unboldmath}
\times \mathbf{a}^{\mu }+g_{\rho }\mbox{\boldmath $\pi$\unboldmath}\times (
\mbox{\boldmath $\pi$\unboldmath}\times \mbox{\boldmath $\rho$\unboldmath}
^{\mu })]_{3}.  \label{Eq.29}
\end{equation}
The strong-interaction Lagrangian can be written in the following form
\begin{equation}
\mathcal{L}_{N\pi \sigma \omega }^{str}=\mathcal{L}_{N\pi \sigma \omega
}^{(0)}+\mathcal{L}_{N\pi \sigma \omega }^{int}+\frac{1}{2}g_{\rho }^{2}v^{2}
\mathbf{a}_{\mu }^{2}-g_{\rho }v\mathbf{a}_{\mu }\partial ^{\mu }
\mbox{\boldmath $\pi$\unboldmath},  \label{Eq.30}
\end{equation}
where the free Lagrangian $\mathcal{L}_{N\pi \sigma \omega }^{(0)}$\ of the
nucleon, pion, sigma and omega reads
\begin{equation}
\mathcal{L}_{N\pi \sigma \omega }^{(0)}=\bar{N}(i\gamma ^{\mu }\partial
_{\mu }-m_{N})N+\frac{1}{2}[(\partial _{\mu }
\mbox{\boldmath
$\pi$\unboldmath})^{2}-m_{\pi }^{2}\mbox{\boldmath $\pi$\unboldmath}^{2}]+
\frac{1}{2}[(\partial _{\mu }\sigma )^{2}-m_{\sigma }^{2}\sigma ^{2}]-\frac{1
}{4}\omega _{\mu \nu }^{2}+\frac{1}{2}m_{\omega }^{2}\omega _{\mu }^{2}.
\label{Eq.31}
\end{equation}
The interaction $\mathcal{L}_{N\pi \sigma \omega }^{int}$ is not needed in
this section, and is given in Eq.(\ref{Eq.A-N-strong}) of Appendix \ref
{App:A}. The third term on the right in Eq.(\ref{Eq.30}) arises due to the
nonzero VEV of the scalar field $\phi $. It gives an additional contribution
to the mass of the $a_{1}$ meson
\begin{equation}  \label{Eq.31bis}
m_{a}^{2}=m_{\rho}^{2}+g_{\rho }^{2}v^{2}=g_{\rho }^{2}(v^{2}+\frac{1}{4}u^{2}).
\end{equation}
The last term in Eq.(\ref{Eq.30}) mixes the pion field with the axial-meson
field.

What remains to be specified are the relations between the masses of the
nucleon, sigma, pion, and the parameters of the potential. They read as
follows
\begin{equation}
m_{N}=g_{\pi }v,\;\ \ \ \ \ \ \ \ \ \ m_{\sigma }^{2}=2\lambda v^{2}+m_{\pi
}^{2},\;\ \ \ \ \ \ \ \ \ \ m_{\pi }^{2}=\frac{c}{v},\;\ \ \ \ \ \ \ \ \ \ \
\mu ^{2}=\frac{1}{2}(m_{\sigma }^{2}-3m_{\pi }^{2}).  \label{Eq.32}
\end{equation}

%%%%%%%%%%%%%%%%%%%%%%%%%%%%%%%%%%%%%%%%%%%%%%%%%%%%%%%%%%%%%%%%%%%%%%%%%%%%
%%%%%%%%%%%%%%%%%%%%%%%%%%%%%%%%%%%%%%%%%%%%%%%%%%%%%%%%%%%%%%%%%%%%%%%%%%%%

\subsection{\label{subsec:Removing} Removing mixing terms in Lagrangian}

The Lagrangians obtained so far are still not complete. They contain
bilinear terms which mix different fields, namely $\mathbf{a}_{\mu }$ and $
\mbox{\boldmath $\pi$\unboldmath},$ $\mathbf{a}_{\mu }$ and $\mathbf{Z},$ $
\mbox{\boldmath $\rho$\unboldmath}_{\mu }$ and $\mathbf{H}.$ To remove these
terms we will follow the method of \cite{Pre99}, with some variations.
Collecting the mixing terms from Eq.(\ref{Eq.19}) and Eq.(\ref{Eq.30}) one
gets
\begin{eqnarray}
\mathcal{L}_{mix} &=&-g_{\rho }v\mathbf{a}_{\mu }\partial ^{\mu }
\mbox{\boldmath
$\pi$\unboldmath}+m_{\rho }(\mathbf{a}_{\mu }\partial ^{\mu }\mathbf{Z}+
\mbox{\boldmath $\rho $\unboldmath}_{\mu }\partial ^{\mu }\mathbf{H}
)=-g_{\rho }\partial ^{\mu }\mathbf{a}_{\mu }\,(\frac{u}{2}\mathbf{Z}-v
\mbox{\boldmath $\pi$\unboldmath})-m_{\rho }\partial ^{\mu }
\mbox{\boldmath$\rho$\unboldmath}_{\mu }\mathbf{H}  \nonumber \\
&=&-m_{a}\partial ^{\mu }\mathbf{a}_{\mu }\mathbf{\tilde{Z}}-m_{\rho
}\partial ^{\mu }\mbox{\boldmath $\rho$\unboldmath}_{\mu }\mathbf{H}\,,
\label{Eq.c33}
\end{eqnarray}
where we dropped a full divergence in the first line and used the following
definition
\begin{equation}
\left(
\begin{array}{c}
\mathbf{\tilde{Z}} \\
\tilde{\mbox{\boldmath $\pi$\unboldmath}}
\end{array}
\right) =\left(
\begin{array}{cc}
\cos \theta _{\pi } & -\sin \theta _{\pi } \\
\sin \theta _{\pi } & \cos \theta _{\pi }
\end{array}
\right) \left(
\begin{array}{c}
\mathbf{Z} \\
\mbox{\boldmath $\pi$\unboldmath}
\end{array}
\right)  \label{Eq.c34}
\end{equation}
of the new fields $\mathbf{\tilde{Z}}$ and $\tilde{
\mbox{\boldmath
$\pi$\unboldmath}}.$ The mixing angle $\theta _{\pi }$\ is determined by $
\tan \theta _{\pi }=2v/u,$ \textit{i.e.} by the ratio of the VEV's of the
scalar fields in the two sectors of the Lagrangian. The transformation (\ref
{Eq.c34}) leaves the sum of the kinetic terms invariant, ($\partial _{\mu }
\mbox{\boldmath $\pi$\unboldmath})^{2}+(\partial _{\mu }\mathbf{Z})^{2}=$($
\partial _{\mu }\tilde{\mbox{\boldmath
$\pi$\unboldmath}})^{2}+(\partial _{\mu }\mathbf{\tilde{Z}})^{2}.$ The
mixing terms in Eq.(\ref{Eq.c33}) can now be removed by adding the
gauge-fixing term $\mathcal{L}_{GF}=-(\partial ^{\mu }\mathbf{a}_{\mu }-\xi
m_{a}\mathbf{\tilde{Z}})^{2}/2\xi -(\partial ^{\mu }
\mbox{\boldmath
$\rho$\unboldmath}_{\mu }-\xi m_{\rho }\mathbf{H})^{2}/2\xi $ similarly to
the procedure fixing the so-called \ $R_{\xi }$ \ gauge in gauge theories
(\cite{Pes95}, Ch.21). For the sum we obtain
\begin{equation}
\mathcal{L}_{mix}+\mathcal{L}_{GF}=-\frac{1}{2\xi }(\partial ^{\mu }\mathbf{a
}_{\mu })^{2}-\frac{1}{2\xi }(\partial ^{\mu }
\mbox{\boldmath
$\rho$\unboldmath}_{\mu })^{2}-\frac{m_{a}^{2}\xi }{2}\mathbf{\tilde{Z}}^{2}-
\frac{m_{\rho }^{2}\xi }{2}\mathbf{H}^{2},  \label{Eq.c35}
\end{equation}
which shows that $\mathbf{\tilde{Z}}$ \ and $\mathbf{H}$\ are fictitious
fields with masses $\ m_{a}\xi ^{1/2}$ and $m_{\rho }\xi ^{1/2}$
respectively. These fields do not contribute to physical processes because
their contribution is always canceled by the $\xi $-dependent part of the $
\mathbf{a}_{\mu }$ or $\mbox{\boldmath
$\rho$\unboldmath}_{\mu }$ propagator \cite{Pes95} (Ch.21.1). We will choose
the unitary gauge \ $\xi \rightarrow \infty $ in which $\mathbf{\tilde{Z}}$
\ and $\mathbf{H}$ completely decouple and the propagator of the vector
meson takes the form\ $\ i(-g^{\mu \nu }+k^{\mu }k^{\nu
}/m^{2})/(k^{2}-m^{2}+i0)$. In this gauge $\mathbf{\tilde{Z}}$
($\mathbf{H}$) provides a longitudinal degree of freedom to the
massive $a_{1}$ ($\rho $) meson \footnote{
Strictly speaking the above arguments apply only in the chiral limit $m_{\pi
}=0$, otherwise the pion mass term in Eq.(\ref{Eq.31}) also contributes to
the mass of the $\tilde{\mathbf{Z}}.$ The pion can be given mass after the
transformation to the new fields is done and the gauge is chosen.}. Setting $
\ \mathbf{\tilde{Z}}=0$ in Eq.(\ref{Eq.c34}) gives $
\mbox{\boldmath
$\pi$\unboldmath}=\cos \theta _{\pi }\tilde{
\mbox{\boldmath
$\pi$\unboldmath}}=(m_{\rho }/m_{a})\tilde{\mbox{\boldmath
$\pi$\unboldmath}}$ and $\mathbf{Z}=\sin \theta _{\pi }\tilde{
\mbox{\boldmath $\pi$\unboldmath}}=(g_{\rho }v/m_{a})\tilde{
\mbox{\boldmath
$\pi$\unboldmath}}$. The latter formulas have been obtained in \cite{Pre99}
in a slightly different way. It is convenient to use the notation $X_{\pi
}\equiv (m_{\rho }/m_{a})^{2}.$ Then in all formulas of the previous
sections we just have to set \ $\mathbf{H}=0$ \ and make the replacements
\begin{equation}
\mbox{\boldmath $\pi$\unboldmath}\rightarrow \sqrt{X_{\pi }}\tilde{
\mbox{\boldmath
$\pi$\unboldmath}}\,,\;\;\ \ \ \ \ \;\;\;\;\;\;\mathbf{Z}\rightarrow \sqrt{
1-X_{\pi }}\tilde{\mbox{\boldmath $\pi$\unboldmath}}\,  \label{Eq.c36}
\end{equation}
in terms of the physical pion field $\tilde{\mbox{\boldmath $\pi$\unboldmath}
} $. It is seen, in particular, that choosing $\mathbf{Z}=0$ from the
beginning leads to a different Lagrangian.

%%%%%%%%%%%%%%%%%%%%%%%%%%%%%%%%%%%%%%%%%%%%%%%%%%%%%%%%%%
%%%%%%%%%%%%%%%%%%%%%%%%%%%%%%%%%%%%%%%%%%%%%%%%%%%%%%%%%%

\subsection{\label{subsec:EM-physical}
Electromagnetic interaction in terms of the physical pion field}

Now we are in a position to write the total EM interaction. For the sake of
brevity we omit from now on the ``tilde'' on the pion field and, after the
substitutions of Eq.(\ref{Eq.c36}) are made, use the notation $
\mbox{\boldmath $\pi$\unboldmath}$ for the physical pion. The current
arising from the Higgs sector reads
\begin{eqnarray}
J_{H}^{\mu } &=&(1-X_{\pi })(\mbox{\boldmath $\pi $\unboldmath}\times
\partial ^{\mu }\mbox{\boldmath $\pi
$\unboldmath})_{3}+\sqrt{1-X_{\pi }}m_{\rho }(
\mbox{\boldmath $\pi
$\unboldmath}\times \mathbf{a}^{\mu })_{3}+\frac{1}{2}g_{\rho }[(1-X_{\pi
})\,\mbox{\boldmath $\pi $\unboldmath}\times (
\mbox{\boldmath $\pi
$\unboldmath}\times \mbox{\boldmath
$\rho$\unboldmath}^{\mu })  \nonumber \\
&&+\sqrt{1-X_{\pi }}\eta \,\mbox{\boldmath $\pi $\unboldmath}\times \mathbf{a
}^{\mu }+\sqrt{1-X_{\pi }}\zeta \,\mbox{\boldmath $\pi $\unboldmath}\times
\mbox{\boldmath $\rho$\unboldmath}^{\mu }]_{3}+(
\mbox{\boldmath
$\rho$\unboldmath}^{\mu \nu }\times \mbox{\boldmath $\rho$\unboldmath}_{\nu
}+\mathbf{a}^{\mu \nu }\times \mathbf{a}_{\nu })_{3},  \label{Eq.c38}
\end{eqnarray}
while the contribution from the $N\pi \sigma \omega $ sector reads
\begin{equation}
J_{N\pi \sigma }^{\mu }=\frac{1}{2}\bar{N}\gamma ^{\mu }(1+\tau
_{3})N+[X_{\pi }\,\mbox{\boldmath $\pi$\unboldmath}\times \partial ^{\mu }
\mbox{\boldmath $\pi$\unboldmath}-g_{\rho }\sqrt{X_{\pi }}(v+\sigma )\,
\mbox{\boldmath $\pi$\unboldmath}\times \mathbf{a}^{\mu }+g_{\rho }X_{\pi }\,
\mbox{\boldmath $\pi$\unboldmath}\times (\mbox{\boldmath $\pi$\unboldmath}
\times \mbox{\boldmath $\rho$\unboldmath}^{\mu })]_{3} .
\label{Eq.c39}
\end{equation}
Adding these currents, and noticing that \ $\sqrt{1-X_{\pi }}m_{\rho
}=g_{\rho }v\sqrt{X_{\pi }},$ we obtain the total EM\ current
\begin{eqnarray}
J^{\mu } &=&\frac{1}{2}\bar{N}\gamma ^{\mu }(1+\tau _{3})N+[
\mbox{\boldmath
$\pi$\unboldmath}\times \partial ^{\mu }\mbox{\boldmath $\pi$\unboldmath}+
\frac{1}{2}g_{\rho }(1+X_{\pi })\,\mbox{\boldmath $\pi$\unboldmath}\times (
\mbox{\boldmath $\pi$\unboldmath}\times \mbox{\boldmath $\rho$\unboldmath}
^{\mu })-g_{\rho }\sqrt{X_{\pi }\,}\sigma \, \mbox{\boldmath $\pi$\unboldmath}
\times \mathbf{a}^{\mu }
\nonumber \\ &&
+\mbox{\boldmath $\rho$\unboldmath}^{\mu \nu }\times\mbox{\boldmath
$\rho$\unboldmath}_{\nu }+\mathbf{a}^{\mu \nu }\times \mathbf{a}_{\nu }+
\frac{1}{2}g_{\rho }\,\sqrt{1-X_{\pi }}\,\eta \,
\mbox{\boldmath $\pi
$\unboldmath}\times \mathbf{a}^{\mu }+\frac{1}{2}g_{\rho }\,\sqrt{1-X_{\pi }}
\,\zeta \,\mbox{\boldmath $\pi$\unboldmath}\times
\mbox{\boldmath
$\rho$\unboldmath}^{\mu }]_{3}.
\label{Eq.c40}
\end{eqnarray}
The nonminimal EM interaction remains the same as in Eq.(\ref{Eq.18}). It
describes the interaction with the intrinsic magnetic moment of the $\rho $
and $a_{1}$ mesons, which is equal to one in this model. The gyromagnetic
ratio for the $\rho $ ($a_{1}$) turns out to be $2$ in units of $\
e/2m_{\rho }$ ($e/2m_{a}$). This is analogous to the nonminimal EM
interaction in GWS theory and in QHD-II \cite{Ser86}.

Eq.(\ref{Eq.c40}) is one of the important results of the paper. It shows the
following features. The pion EM current is restored to its original form
(the current of the free pion). Due to a cancellation between the currents,
the term proportional to ($\mbox{\boldmath $\pi $\unboldmath}\times \mathbf{a
}^{\mu })$ disappears. Therefore there is no $a\pi \gamma $ interaction on
the tree level. As a result of the diagonalization in Eq.(\ref{Eq.11}) the $
\rho $ meson does not couple directly to the photon, so there is no explicit
vector-meson dominance of the EM interaction. The EM Lagrangian includes the
3-field interactions $\gamma NN,\gamma \pi \pi ,\gamma \rho \rho ,\gamma aa,$
as well as the 4-field interaction pieces $\gamma \pi \pi \rho ,\gamma \pi
\sigma a,\gamma \rho \rho \rho ,\gamma \rho aa,\gamma \pi a\eta $ and $
\gamma \pi \rho \zeta .$ The latter vertices are important for the EM gauge
invariance of the amplitudes which will be calculated in Sect.~\ref
{sec:Meson decays}.

For completeness we present the strong-interaction Lagrangian which follows
from Eqs.(\ref{Eq.19}) and (\ref{Eq.30}),
\begin{equation}
\mathcal{L}^{str}=\mathcal{L}_{\eta \zeta }^{(0)}+\mathcal{L}_{\rho a}^{(0)}+
\mathcal{L}_{N\pi \sigma \omega }^{(0)}+\mathcal{L}_{N\pi \sigma \omega
}^{int}+\mathcal{L}_{H}^{int}\, ,
\label{Eq.c41}
\end{equation}
\begin{eqnarray}
\mathcal{L}_{N\pi \sigma \omega }^{int} &=&-\bar{N} [g_{\pi }(\sigma +i\sqrt{
X_{\pi }}\gamma _{5} \mbox{\boldmath $\tau$\unboldmath}
\mbox{\boldmath
$\pi$\unboldmath})+g_{\rho }\gamma ^{\mu } \frac{
\mbox{\boldmath
$\tau$\unboldmath}}{2}(\mbox{\boldmath $\rho$\unboldmath}_{\mu } +\gamma _{5}
\mathbf{a}_{\mu })+g_{\omega }\gamma ^{\mu }\omega _{\mu }]N
\nonumber \\
&&-\lambda (\sigma ^{2}+X_{\pi }\mbox{\boldmath $\pi$\unboldmath}
^{2})[v\sigma +\frac{1}{4}(\sigma ^{2} +X_{\pi }
\mbox{\boldmath$\pi$\unboldmath}^{2})] -\frac{1}{2}g_{\rho }(1+X_{\pi })
\mbox{\boldmath
$\rho$\unboldmath}_{\mu }(\mbox{\boldmath $\pi$\unboldmath}\times \partial
^{\mu }\mbox{\boldmath $\pi$\unboldmath})  \nonumber \\
&& +g_{\rho }\sqrt{X_{\pi }}\mathbf{a}_{\mu }(
\mbox{\boldmath
$\pi$\unboldmath}\partial ^{\mu }\sigma -\sigma \partial ^{\mu }
\mbox{\boldmath $\pi$\unboldmath}) +\frac{1}{2}g_{\rho }^{2}[X_{\pi }(
\mbox{\boldmath $\pi$\unboldmath} \mathbf{a}_{\mu })^{2}+(\sqrt{X_{\pi }}
\mbox{\boldmath $\pi$\unboldmath} \times \mbox{\boldmath $\rho$\unboldmath}
_{\mu } -\sigma \mathbf{a}_{\mu})^{2}  \nonumber \\
&& -2v\mathbf{a}_{\mu }(\sqrt{X_{\pi }}\mbox{\boldmath $\pi$\unboldmath}
\times \mbox{\boldmath $\rho$\unboldmath}^{\mu }-\sigma \mathbf{a}^{\mu })],
\label{Eq.c42}
\end{eqnarray}
\begin{eqnarray}
\mathcal{L}_{H}^{int} &=&\frac{1}{8}g_{\rho }^{2}\{(
\mbox{\boldmath
$\rho$\unboldmath}_{\mu }^{2}+\mathbf{a}_{\mu }^{2})[\eta ^{2}+\zeta
^{2}+2u\eta +(1-X_{\pi })\mbox{\boldmath $\pi$\unboldmath}^{2}]+4
\mbox{\boldmath
$\rho$\unboldmath}_{\mu }\mathbf{a}^{\mu }(u+\eta )\zeta \}  \nonumber \\
&&+\frac{1}{2}g_{\rho }\sqrt{1-X_{\pi }}[\mbox{\boldmath $\rho$\unboldmath}
_{\mu }(\zeta \partial ^{\mu }\mbox{\boldmath $\pi $\unboldmath}-
\mbox{\boldmath $\pi $\unboldmath}\partial ^{\mu }\zeta )+\mathbf{a}_{\mu
}(\eta \partial ^{\mu }\mbox{\boldmath $\pi $\unboldmath}-
\mbox{\boldmath$\pi $\unboldmath}\partial ^{\mu }\eta )]  \nonumber \\
&& -\frac{\lambda _{H}}{32}[\eta ^{4}+\zeta ^{4}+4u\eta ^{3}+6\eta \zeta
^{2}(2u+\eta ) +(1-X_{\pi })^{2}\mbox{\boldmath $\pi $\unboldmath}^{4}
\nonumber \\
&& +2(1-X_{\pi })\mbox{\boldmath $\pi$\unboldmath}^{2} (\eta ^{2}+\zeta
^{2}+2u\eta )].
\label{Eq.c43}
\end{eqnarray}
The Lagrangian $\mathcal{L}_{\rho a}^{(0)}$ is given in Eq.(\ref{Eq.21}), where
now we have to take the mass of the $a_{1}$ meson from Eq.(\ref{Eq.31bis}).

Although the EM current and strong-interaction Lagrangian look somewhat
complicated, they contain only simple vertices with at most one derivative.
This greatly simplifies practical calculations. It is seen from Eq.(\ref
{Eq.c42}) that, apart from the $\rho \pi \pi $ coupling, the strength of the
coupling to the pion is scaled down by a factor $\sqrt{X_{\pi }}$ for each
pion field operator. At the same time the coupling to the pion from the
Higgs Lagrangian in Eq.(\ref{Eq.c43}) is scaled by a factor $\sqrt{1-X_{\pi }}$.
 It also follows that the
$\rho$-meson coupling to the pion, $g_{\rho \pi \pi }=g_{\rho }(1+X_{\pi })/2$, is
not equal to the $\rho$ coupling to the nucleon, $g_{\rho NN}=g_{\rho }$. So
in this model the $\rho $ does not couple universally to the hadrons.

The presence of the Higgs fields $\eta \ $ and $\zeta $ may seem as an
obstacle. However, as was argued in ref.~\cite{Ser92}, these fields serve as
regulators in the calculation of loop integrals. By taking the mass $m_{H}$
very large the Higgs contributions can be suppressed in many cases. We will
study this aspect while calculating meson decays and low-energy pion-pion
scattering in the next sections.

It is interesting to note that the EM current in Eq.(\ref{Eq.c40}) can be
derived in a simpler way, directly from the strong-interaction Lagrangian in
Eq.(\ref{Eq.c41}). Indeed, the minimal substitution $\partial _{\mu
}\rightarrow \partial _{\mu }+ie\hat{Q}A_{\mu }$ ($\hat{Q}$ is the charge
operator) applied to all charged fields leads to Eq.(\ref{Eq.c40}). At the
same time the nonminimal EM interaction in Eq.(\ref{Eq.18}) cannot be
obtained in this way. The EM current satisfies the relation
\begin{equation}
J^{\mu }=I_{3}^{\mu }+\frac{1}{2}B^{\mu },  \label{Eq.c44}
\end{equation}
where $I_{3}^{\mu }$ is the 3d component of the conserved isospin current $
\mathbf{I}^{\mu }$, and $B^{\mu }$ is the conserved baryon current. The
latter is $B^{\mu }=\bar{N}\gamma ^{\mu }N$, while the expression for the
former is given by
\begin{eqnarray}
\mathbf{I}^{\mu } &=& \frac{1}{2}\bar{N}\gamma ^{\mu }
\mbox{\boldmath
$\tau$\unboldmath}N+\mbox{\boldmath $\pi$\unboldmath}\times \partial ^{\mu }
\mbox{\boldmath $\pi$\unboldmath}+\frac{1}{2}g_{\rho }(1+X_{\pi })\,
\mbox{\boldmath $\pi$\unboldmath}\times (\mbox{\boldmath $\pi$\unboldmath}
\times \mbox{\boldmath $\rho$\unboldmath}^{\mu })-g_{\rho }\sqrt{X_{\pi }\,}
\sigma \mbox{\boldmath $\pi$\unboldmath}\times \mathbf{a}^{\mu }  \nonumber
\\
&&+\mbox{\boldmath $\rho$\unboldmath}^{\mu \nu }\times
\mbox{\boldmath
$\rho$\unboldmath}_{\nu }+\mathbf{a}^{\mu \nu }\times \mathbf{a}_{\nu }+
\frac{1}{2}g_{\rho }\,\sqrt{1-X_{\pi }}\,\eta \,
\mbox{\boldmath $\pi
$\unboldmath}\times \mathbf{a}^{\mu }+\frac{1}{2}g_{\rho }\,\sqrt{1-X_{\pi }}
\,\zeta \,\mbox{\boldmath
$\pi $\unboldmath}\times \mbox{\boldmath $\rho$\unboldmath}^{\mu }.
\label{Eq.c45}
\end{eqnarray}

Conservation of the isospin current is a consequence of the symmetry of
the strong-interaction Lagrangian in Eqs.(\ref{Eq.c41}-\ref{Eq.c43}) with
respect to the \textit{global} $SU(2)_{V}$ \ isospin transformations.
Indeed, it can be readily verified that $\mathbf{I}^{\mu }$ in Eq.(\ref
{Eq.c45}) is the corresponding Noether current. It also follows
from Eqs.(\ref{Eq.c41}-\ref{Eq.c43}) that the $\rho $ meson is
coupled to a source current $\ \mathbf{J}_{\rho }^{\mu }
=-g_{\rho }^{-1} (\delta /\delta \mbox{\boldmath
$\rho$\unboldmath}_{\mu } ) \,(\mathcal{L}_{N\pi \sigma \omega }^{int}+
\mathcal{L}_{H}^{int}),$ which is in general different from $\mathbf{I}^{\mu
}.$ The source current is closely related to a current corresponding to the
underlying \textit{local} $\ SU(2)_{V}$ \ symmetry, which is ``hidden''.
This symmetry may be classified according to ref.~\cite{Sal72} as a symmetry
of the 2nd kind. A more detailed study of this aspect is beyond the scope of
the present paper. For a related discussion in QHD-II see Appendix D of
ref.~ \cite{Ser86}.

%%%%%%%%%%%%%%%%%%%%%%%%%%%%%%%%%%%%%%%%%%%%%%%%%%%%%%%%%%%%%%%%%%%%%%%%%%%%
%%%%%%%%%%%%%%%%%%%%%%%%%%%%%%%%%%%%%%%%%%%%%%%%%%%%%%%%%%%%%%%%%%%%%%%%%%%%

\section{\label{sec:Meson decays} Decay of the vector and axial-vector mesons}

First we consider the decay of the mesons which can be obtained on the tree
level. These are the decays $\rho \rightarrow \pi \pi ,\,\ \rho \rightarrow
\pi \pi \gamma ,$ $a_{1}\rightarrow \pi \rho ,\,\ a_{1}\rightarrow \pi
\sigma ,$ and $\sigma \rightarrow \pi \pi .$ We will need the general
expression for the width of the decay $\ A(Q)\rightarrow B(p)+C(q)$\ \ in
the rest frame of the decaying particle with the mass $M_{A}$ and spin
$J_{A} $
\begin{equation}
\Gamma =\frac{|\mathbf{p}|}{8\pi M_{A}^{2}}\frac{1}{(2J_{A}+1)}\sum_{spins}|
\mathcal{M}|^{2},  \label{Eq.46}
\end{equation}
where $Q,p$ and $q$ are the corresponding 4-momenta, such that \ $Q=p+q.$
The 3-momentum of the particles in the final state is $\mathbf{p}$
($\mathbf{q=-p}$) and
the sum runs over the polarizations of all particles. The decay width for
the $\rho\rightarrow\pi\pi\gamma$  will be discussed below.

We first fix the parameters of the model. In general, the coupling constants
($g_{\pi },g_{\rho ,}g_{\omega }),$ the parameters of the potentials ($\mu
,\lambda ,c,\mu _{H},\lambda _{H}),$ and all the masses can be considered as
parameters. There are however many relations between these parameters:
Eq.(\ref{Eq.22}), Eq.(\ref{Eq.31bis}) and Eq.(\ref{Eq.32}). Simple considerations
show that if we choose the masses of the nucleon, pion, rho, $a_{1},$ and
omega equal to their experimental values, then there remain only four free
parameters: $g_{\rho ,}g_{\omega }, $ the poorly known sigma mass $m_{\sigma
},$ and the unknown mass \ $m_{H} $ of the Higgs particles. We will fix the
coupling $g_{\rho }$ from the $\rho \rightarrow \pi \pi $ decay width. It is
seen from Eq.(\ref{Eq.c42}) that the \ $\rho \rightarrow \pi \pi $ decay is
determined by the matrix element $\mathcal{M=}\ g_{\rho \pi \pi }\,\epsilon
(\rho )\cdot (q-p)\varepsilon ^{ijk}$. The polarization vector of the $\rho $
is denoted by $\epsilon (\rho )$, and Latin indices label the charge states
of the mesons. From the experimental width 150.2 MeV one finds $g_{\rho \pi
\pi }=6.04.$ Taking $m_{a}=$1.23 GeV \cite{PDG}, we obtain $g_{\rho }=8.68,$
$g_{\pi }=$ 8.49, $v=111$ MeV, \ $u/2=88.7$ MeV, and $X_{\pi }=0.392.$
Curiously enough, the ratio $g_{\rho }/g_{\rho \pi \pi }=2/(1+X_{\pi })$
appears to be 1.437, which is close to a factor $\sqrt{2}$ (with a
deviation of 1.6\%). It follows that $X_{\pi }\approx \sqrt{2}-1$ and $m_{a}
\approx (1+\sqrt{2})^{1/2} m_{\rho }.$

%%%%%%%%%%%%%%%%%%%%%%%%%%%%%%%%%%%%%%%%%%%%%%%%%%%%%%%%%%%%%%%%%%%%

\subsection{\label{subsec:Strong decays}Strong decays}

The decay $a_{1}\rightarrow \pi \rho $ is governed by the matrix element $\
\mathcal{M=}-ig_{\rho }^{2}v\sqrt{X_{\pi }}\,\epsilon (a)\cdot \epsilon
^{\ast }(\rho )\varepsilon ^{ijk},$ where $\epsilon (a)$ [$\epsilon (\rho )
$] is the polarization vector of the $a_{1}$ $(\rho )$ meson. The $a_{1} \pi
\rho $ vertex is simpler than that in GLSM \cite{Gas69,Ko94}, or in the
''massive '' Yang-Mills approach \cite{Mei88}. Moreover it does not vanish
for any invariant mass of the $a_{1.}$ The calculated width of 272 MeV can
be compared with the experimental estimate 150 to 361 MeV~\cite{PDG}. In
general, this decay is characterized by the two amplitudes, $F$ and $G,$\
defined through \ $\mathcal{M}=F\,\epsilon (a)\cdot \epsilon ^{\ast }(\rho
)+G\,\epsilon (a)\cdot p\;\epsilon ^{\ast }(\rho )\cdot Q$. Those in turn
define the S- and D-wave amplitudes \cite{Isg89}
\begin{equation}
F_{S}=\frac{\sqrt{4\pi }}{3m_{\rho }}[(\varepsilon _{\rho }+2m_{\rho })F
+\mathbf{p}^{2}m_{a}G],\;\ \ \ \ \ \ \ \ F_{D}=-\frac{\sqrt{8\pi }}{3m_{\rho }
}[(\varepsilon _{\rho }-m_{\rho })F+\mathbf{p}^{2}m_{a}G]\,,  \label{Eq.47}
\end{equation}
where $\varepsilon _{\rho }$ is the energy of the $\rho $ meson in the final
state$.$ Since $G=0$ in the present model we obtain the D/S ratio
$F_{D}/F_{S}=-4.62$\%, which agrees in sign and order of magnitude with the
experimental ratio $-10.7\pm$1.6\% \cite{PDG}.

The $a_{1}\rightarrow \pi \sigma $ matrix element, according to Eq.(\ref
{Eq.c42}), is $\ \mathcal{M=}\ g_{\rho }\sqrt{X_{\pi }}\epsilon (a)\cdot
(q-p)\delta ^{ik}.$ The corresponding calculated width comes out 46 MeV and
the branching ratio is $\Gamma (a_{1}\rightarrow \pi \sigma )/\Gamma
_{tot}\approx \ \Gamma (a_{1}\rightarrow \pi \sigma )/[\Gamma
(a_{1}\rightarrow \pi \rho )+\Gamma (a_{1}\rightarrow \pi \sigma )] \approx$
14\%. From the branching ratio given by PDG we can estimate the
corresponding width as 32 to 147 MeV. Of course this process is not very
well defined, in view of the uncertain status of the $\sigma $ meson. We
used the value 770 MeV for the mass of the sigma.

Next we address the issue of the width of the $\sigma $ meson.
This subject has been discussed extensively in~\cite{Ko94}, mainly because
in the linear $\sigma $ model the width is too large, and sometimes larger
than its mass, which makes it difficult to identify the $\sigma $ with a
particle state. In our model the matrix element of the $\sigma \rightarrow
\pi \pi $ decay is given by \ $\mathcal{M=}-2i\lambda vX_{\pi }\,\delta
^{ij}.$ Calculation yields $\Gamma =149$ MeV for $m_{\sigma }=770$ MeV$,$
and $\Gamma =346$ MeV for $m_{\sigma }=1$ GeV.\ If we had used - as is
appropriate in the linear $\sigma $ model - the values $X_{\pi }=1$ and
$v=f_{\pi }$, where $f_{\pi }=93.2$ MeV is the pion weak-decay constant, then
we would indeed have obtained a very large width of $1.391$ GeV ($3.227$ GeV)
for $m_{\sigma }=770 $ MeV\ ($m_{\sigma }=1$ GeV). The width, however, is reduced
considerably due to the factor $X_{\pi }$ in the $\sigma \pi \pi $ vertex,
and to a lesser extent due to the difference between $v$ and $f_{\pi }.$ We
should also mention that the vertex, because of its simple structure, does
not vanish for any values of the invariant mass of the $\sigma .$ The
vanishing of the vertex is an undesirable feature of GLSM, as was pointed
out in ref.~\cite{Urb01}.

%%%%%%%%%%%%%%%%%%%%%%%%%%%%%%%%%%%%%%%%%%%%%%%%%%%%%%%%%%%%%%%%

\subsection{\label{subsec:EM decays} Electromagnetic decays}

\subsubsection{\label{subsubsec:rho-pi-pi-gam}
$\protect\rho ^{0}\rightarrow \protect\pi ^{+}\protect\pi ^{-}
\protect\gamma $ decay}

\begin{figure}[tbp]
\includegraphics[width=15cm]{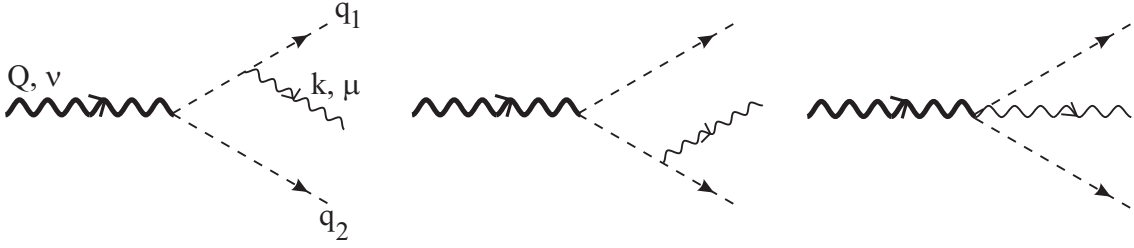}
\caption[QHD]{\label{fig:1}
Diagrams for the $\protect\rho^0 \rightarrow \protect\pi^{+}
\protect\pi^{-} \protect\gamma$ decay.}
\end{figure}

Let us consider the EM decay $\rho ^{0}(Q)\rightarrow \pi ^{+}(q_{1})\,+\
\pi ^{-}(q_{2})+\,\ \gamma (k)$, which can be described by the tree-level
amplitude shown on Fig.~\ref{fig:1}. The matrix element can be written,
using Eq.(\ref{Eq.c40}) and Eq.(\ref{Eq.c42}), as \ $\mathcal{M}=\epsilon
_{\nu }(\rho )\epsilon _{\mu }^{\ast }(\gamma )M^{\mu \nu }$ where
\begin{equation}
M^{\mu \nu }=eg_{\rho \pi \pi }\{\frac{(2q_{2}+k)^{\mu }(2q_{1}-Q)^{\nu }}{
2k\cdot q_{2}}+\frac{(2q_{1}+k)^{\mu }(2q_{2}-Q)^{\nu }}{2k\cdot q_{1}}
+2g^{\mu \nu }\}.  \label{Eq.48}
\end{equation}
The last term comes from the $\gamma \pi \pi \rho $ vertex in Eq.(\ref
{Eq.c40}). The total amplitude is gauge invariant, $\ k_{\mu }M^{\mu \nu
}=0. $ Calculation of the decay width involves integration over invariant
masses of the pairs of particles in the final state,
\begin{equation}
\Gamma =\frac{1}{(2\pi )^{3}32m_{\rho }^{3}}\int_{4m_{\pi }^{2}}^{m_{\rho
}^{2}}\mbox{\rm d}m_{\pi \pi }^{2}\int_{m_{\min }^{2}}^{m_{\max }^{2}}
\mbox{\rm d}m_{\gamma \pi }^{2}\, \frac{1}{3}\sum_{spins}|\mathcal{M}|^{2}.
\label{Eq.49}
\end{equation}
The limits in the integral over \ $m_{\gamma \pi }^{2}$ are $\ m_{\max/\min
}^{2}=m_{\pi }^{2}+2k^{\ast }(E^{\ast } \pm q^{\ast }),$
where \ $k^{\ast },E^{\ast }$ and $q^{\ast }$ are respectively
the photon momentum, pion energy and pion
momentum in the rest frame of the $\pi -\pi $ system,
\begin{equation}
k^{\ast }=\frac{m_{\rho }^{2}-m_{\pi \pi }^{2}}{2m_{\pi \pi }},\;\ \ \ \ \ \
\ \ \ \ \ \ E^{\ast }=\frac{1}{2}m_{\pi \pi },\;\ \ \ \ \ \ \ \ \ \ \ \
q^{\ast }=(E^{\ast 2}-m_{\pi }^{2})^{1/2}\,.  \label{Eq.50}
\end{equation}
The sum and average over polarizations is most easily evaluated in the
system where the OZ axis is along the photon 3-momentum $\mathbf{k}$. We
obtain
\begin{eqnarray}
\frac{1}{3}\sum_{spins}|\mathcal{M}|^{2} &=&\frac{4}{3} e^{2} g_{\rho \pi
\pi}^{2} \left\{ 2+\frac{\mathbf{q}_{2}^{2}\mathbf{q}_{1T}^{2}}{(k\cdot q_{1})^{2}
}+\frac{\mathbf{q}_{1}^{2}\mathbf{q}_{2T}^{2}}{(k\cdot q_{2})^{2}}
\right. \nonumber \\
&& \left. +2\mathbf{q}_{1T}\cdot\mathbf{q}_{2T}\left[ \frac{\mathbf{q}_{1}\cdot
\mathbf{q}_{2}
}{(k\cdot q_{1})(k\cdot q_{2})}-\frac{1}{k\cdot q_{1}}-\frac{1}{k\cdot q_{2}}
\right] \right\} ,  \label{Eq.51}
\end{eqnarray}
where $\ \mathbf{q}_{1T},\mathbf{q}_{2T}\ $ are the components of the pion
momenta orthogonal to the OZ axis, \textit{e.g.} $\mathbf{q}_{1T}
=\mathbf{q}_{1}-\mathbf{k}(\mathbf{q}_{1}\cdot\mathbf{k})/\mathbf{k}^{2}$.

The decay $\rho ^{0}\rightarrow \pi ^{+}\pi ^{-}\gamma $ includes the
bremsstrahlung from the charged pions and is infrared divergent at small
photon energies. Experimentally a cut-off is introduced while measuring the
decay width, namely $k>k_{\min }.$ \ This means that in the integral in
Eq.(\ref{Eq.49}) the invariant mass (squared) of the $\pi -\pi $ pair has an
upper limit \ $m_{\rho }^{2}-2m_{\rho }k_{\min }$. \ The value of the
integral in Eq.(\ref{Eq.49}) depends strongly on $k_{\min }.$ For $k_{\min
}=50$ MeV we obtain $\Gamma =1.73$ MeV, while for $k_{\min }=60$ MeV \ we
get $\ \Gamma =1.49$ MeV. The PDG review \cite{PDG} gives the value 1.487$%
\pm0.240$ MeV for the photon energies above 50 MeV.

%%%%%%%%%%%%%%%%%%%%%%%%%%%%%%%%%%%%%%%%%%%%%%%%%%%%
%%%%%%%%%%%%%%%%%%%%%%%%%%%%%%%%%%%%%%%%%%%%%%%%%%%%%

\subsubsection{\label{subsubsec:a-pi-gam} $a_{1}^{+}\rightarrow \protect\pi
^{+}\protect\gamma $ decay}

\begin{figure}[tbp]
\includegraphics[width=15cm]{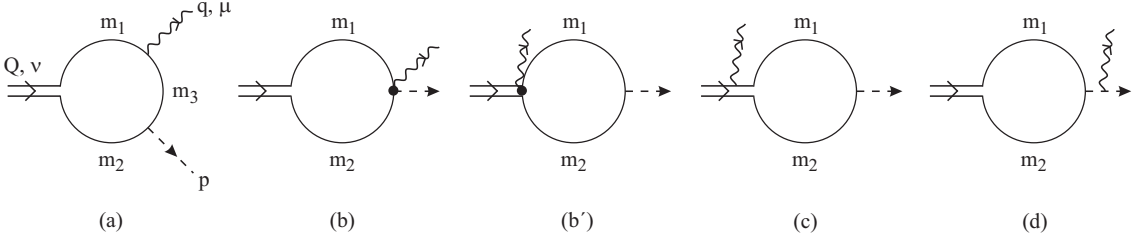}
\caption[QHD]{\label{fig:2}
Diagrams for $a^+ \rightarrow \protect\pi^{+}
\protect\gamma$ decay at the one-loop level. The intermediate propagators
labeled $m_1$, $m_2$, $m_3$ refer to particles as defined in
Table~\protect\ref{tab:diagrams}. Depending on the intermediate state $j$, the
contact diagrams have the structure $(b)$ for $j=1,2,4,6,7$
and the structure $(b^{\prime})$ for
$j=3,5,9,10$. For $j=8$ there is no contact diagram.}
\end{figure}

Next we consider the EM decay $a_{1}^{+}(Q)\rightarrow \pi ^{+}(p)+\,\gamma
(q).$ \ There is no contribution to the matrix element on the tree level,
and we have to include at least the one-loop processes shown in Fig.~\ref
{fig:2} (see also Table~\ref{tab:diagrams}).

\begin{table}[tbp]
\caption{\label{tab:diagrams}
Intermediate particles in the loop diagrams indicated in Fig.~\protect\ref{fig:2}. }
%\begin{center}
\begin{tabular}{c|cccccccccc}
$j$ & 1 &2& 3&4 &5 &6 &7 &8 &9 &10   \\
\hline
$m_1$ & $\pi$ & $\rho$ & $\pi$ & $a$ & $\pi$ & $a$ & $\rho$ & $N$ & $\rho$ & $a $\\
$m_2$ & $\rho$ & $\pi$ & $\sigma$ & $\sigma$ & $\eta$ & $\eta$ & $\zeta$ & $N$
& $a$ & $\rho$ \\
$m_3$ & $\pi$ & $\rho$ & $\pi$ & $a$ & $\pi$ & $a$ & $\rho$ & $N$ & $\rho$ & $a $\\
\end{tabular}
%\end{center}
\end{table}

These contributions can be obtained by attaching the photon line to
the lines of the charged particles in the mixed self-energy operator $\Sigma
_{a\pi }(Q),$ and by adding diagrams coming from the contact terms in the
current (\ref{Eq.c40}). Each one-loop diagram for $\Sigma _{a\pi }(Q)$ gives
rise to four diagrams [labeled (a),(b),(c),(d) in Fig.~\ref{fig:2}] for the EM
process. The amplitude can be written as a sum
\begin{equation}
\mathcal{M=}\sum_{j=1}^{10}\mathcal{M}_{j}=\sum_{j=1}^{10}\epsilon _{\nu
}(a)\epsilon _{\mu }^{\ast }(\gamma )\ M_{j}^{\mu \nu },
\label{Eq.52}
\end{equation}
where $j$ labels the intermediate state in the loop, namely $j=(\pi
^{+}\rho ^{0})$, $(\pi ^{0}\rho ^{+})$, $(\pi ^{+}\sigma )$, $(a^{+} \sigma)$,
$(\pi ^{+} \eta)$, $(a^{+} \eta)$, $(\rho ^{+} \zeta)$, $(p \bar{n})$,
$(\rho^{+} a^{0})$,
and $(\rho^0 a^{+})$. Note that in the present study, because of the
technical complexity, the diagrams with $j=9$ and $j=10$ (containing at
least two vector or axial-vector propagators in the loop) are not included.
Calculation of the amplitudes is cumbersome and we refer to Appendix~\ref
{App:B} for details. However, some features of the calculation are worth
mentioning here.

i) Each amplitude $M_{j}^{\mu \nu }$ is gauge invariant and has the
structure
\begin{equation}
M_{j}^{\mu \nu }=(g^{\mu \nu }-\frac{Q^{\mu }q^{\nu }}{Q\cdot q})\,T_{j},
\label{Eq.53}
\end{equation}
where the $T_j$ are Lorentz scalars.
This serves as an important check of the calculation. The contact
terms $\gamma \pi \pi \rho ,\gamma \pi \sigma a, \gamma\rho a a,
\gamma \pi a \eta $ \ and $ \gamma
\pi \rho \zeta $ in Eq.(\ref{Eq.c40}) are crucial to ensure this property.

ii) The diagrams with the bremsstrahlung from the final pion [labeled (d) in
Fig.~\ref{fig:2}] do not contribute to the matrix element for the on-mass-shell
$a_{1}$ meson. Indeed, the polarization vector of the $a_{1}$ meson
satisfies the relation \ $\epsilon (a)\cdot Q=0.$ \ It is easy to check that
the matrix element corresponding to these diagrams is in each case
proportional to $Q^{\nu },$ and therefore vanishes when multiplied by
$\epsilon _{\nu }(a).$

iii) Some of the diagrams on Fig.~\ref{fig:2} are divergent. However, the
divergent terms from the different diagrams in any $M_{j}^{\mu \nu }$ cancel,
and the total amplitude is finite. The calculations are done in the unitary
gauge using the method of dimensional regularization (see Appendix~\ref
{App:B}), in which the cancellation of divergencies is explicitly verified.

iv) The diagrams with $j=5,6,7$ in Fig.~\ref{fig:2} contain
intermediate Higgs mesons $\eta $ and $\zeta $. If one takes the
mass $m_{H}$ of these mesons very large the amplitudes remain finite and
correspond to a contact-like $\gamma \pi a $ Lagrangian
\begin{equation}
\mathcal{L}_{cont}(\gamma \pi a)=h(\partial _{\mu }A_{\nu }-\partial _{\nu
}A_{\mu })(\mbox{\boldmath $\pi$\unboldmath}\times \partial ^{\mu }\mathbf{a}
^{\nu })_{3}
\label{Eq.54}
\end{equation}
where \ $h=-(4\pi )^{-2}eg_{\rho }^{2}(1-X_{\pi
})^{1/2}/(3 m_{\rho }).$

v) The amplitudes $M_{j}^{\mu \nu }$ for $j=1,2,3$ have both real and
imaginary parts, because the masses of the intermediate particles in these
diagrams satisfy the condition $m_{1}+m_{2}<m_{a}$. \ The amplitudes \
$M_{j}^{\mu \nu }$ for $j=4,...,10$ are real.

vi) The most complicated diagrams are those involving the EM vertex of the
$\rho $ and $a_{1}$ mesons. The vertex for the $\gamma \rho \rho $ (or
$\gamma aa)$ has the form
\begin{eqnarray}
\Gamma ^{\mu \nu \lambda}(q,r,p) &=&e\{g^{\lambda \nu }(r-p)^{\mu }-g^{\mu
\nu }r^{\lambda }+g^{\lambda \mu }p^{\nu }+(g^{\mu \nu }q^{\lambda }-g^{\mu
\lambda }q^{\nu })\}\delta ^{ij}  \nonumber \\
&=&e\{g^{\lambda \nu }(r-p)^{\mu }+g^{\mu \nu }(q-r)^{\lambda }+g^{\lambda
\mu }(p-q)^{\nu }\}\delta ^{ij},
\label{Eq.55}
\end{eqnarray}
where $q$ is the momentum of the photon (with the Lorentz index $\mu ),$ $p$ and $r$
are the momenta of the $\rho $ (with the Lorentz indices $\lambda $ and $\nu$),
and $p+q+r=0$. In the first line we explicitly separated
the minimal EM interaction and coupling to the intrinsic magnetic moment of
the mesons.

vii) In the rest frame of the meson $a_{1},$ where $Q^{\nu }=(m_{a,}\mathbf{0
}),$ one can use the additional relation $Q\cdot \epsilon (\gamma )=0$, as
the photon polarization vectors have only space-like components. Therefore
in the general structure of the amplitude of Eq.(\ref{Eq.53}), only the term
proportional to $g^{\mu \nu }$ contributes.

The width of the $a_{1}^{+}\rightarrow \pi ^{+}\gamma $ decay is
expressed in terms of the $T_{j}$ as follows
\begin{equation}
\Gamma =\frac{|\mathbf{p}|}{8\pi m_{a}^{2}}\frac{2}{3}\,\left|
\sum_{j=1}^{10}T_{j}\right| ^{2}.
 \label{Eq.69}
\end{equation}
The results of the calculation are presented in Table~\ref{tab:a-pi-gam}.

\begin{table}[tbp]
\caption{\label{tab:a-pi-gam}
Contribution of different diagrams on Fig.~\ref{fig:2} to the $a^+
\to \protect\pi^+ \protect\gamma$ decay width. Values are given in KeV. }
%\begin{center} %%%\begin{ruledtabular}
\begin{tabular}{|l|c|c|c|}
\hline
Intermediate state in the diagrams & $m_{a}=1.23$ GeV & $m_{a}=1.089$ GeV &
Experiment \cite{PDG} \\ \hline
$\pi ^{+}\rho ^{0}$ & 469 & 330 &  \\
$\pi ^{+}\rho ^{0}+\pi ^{0}\rho ^{+}$ & 187 & 89 &  \\
$\pi ^{+}\rho ^{0}+\pi ^{0}\rho ^{+}+\pi ^{+}\sigma $ & 211 & 104 &  \\
$\pi ^{+}\rho ^{0}+\pi ^{0}\rho ^{+}+\pi ^{+}\sigma +a^{+}\sigma $ & 845 &
434 &  \\
$\pi ^{+}\rho ^{0}+\pi ^{0}\rho ^{+}+\pi ^{+}\sigma +a^{+}\sigma $+ &  &  &
\\
$\;\;\;\;\;\;\;\;\;\;\;\; + \ [\pi^{+}\eta +a^{+}\eta +\rho ^{+}\zeta ]$ &
646 & 345 &  \\
$\pi ^{+}\rho ^{0}+\pi ^{0}\rho ^{+}+\pi ^{+}\sigma +a^{+}\sigma+ $ &  &  &  \\
$\;\;\;\;\;\;\;\;\;\;\;\; + \ [\pi ^{+}\eta +a^{+}\eta +\rho ^{+}\zeta ]
+p\bar{n}$ & 412 & 192 & 640$\pm 246$ \\ \hline
\end{tabular}
%%%\end{ruledtabular}
%\end{center}
\end{table}

Calculations were performed with two values of the $a_{1}$-meson mass: 1.23
GeV \cite{PDG}, and $\sqrt{2}m_{\rho }=1.089$ GeV. The latter value is often
discussed in the literature \cite{Gil68,Mei88,Wei90}. One notices from
Table~\ref{tab:a-pi-gam} that the different amplitudes strongly interfere. For
example, the $\pi ^{+}\rho ^{0}$ and $\pi ^{0}\rho ^{+}$ amplitudes almost cancel
each other. A substantial contribution comes from the $a^{+}\sigma $ loop
($j=4)$, due to the large value of the constant in front of the integral.
There is a dependence on the $\sigma $ mass, but it is weak. We
used here $m_{\sigma }=770$ MeV. The diagrams in Fig.~\ref{fig:2} containing
the intermediate Higgs mesons give a relatively small contribution, as can
be seen from the sixth row in Table~\ref{tab:a-pi-gam}.
The $N\bar{N}$ diagrams ($j=8)$ by themselves would give a small contribution.
However, due to the
interference with other diagrams, their effect becomes sizeable. On the
whole, the calculation with $m_{a}=1.23$ GeV yields a width in agreement
with experiment. Note however that the diagrams with $j=9$ and $j=10$ were not
included.

%%%%%%%%%%%%%%%%%%%%%%%%%%%%%%%%%%%%%%%%%%%%
%%%%%%%%%%%%%%%%%%%%%%%%%%%%%%%%%%%%%%%%%%%%

\subsubsection{\label{subsubsec:rho-pi-gam} $\protect\rho ^{+}\rightarrow
\protect\pi ^{+}\protect\gamma $ \ decay}

\begin{figure}[tbp]
\includegraphics[width=7cm]{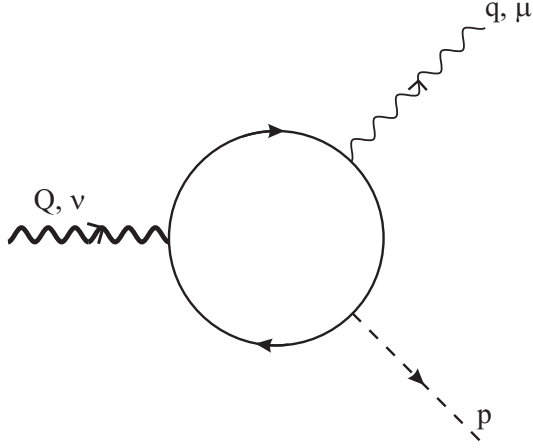}
\caption[QHD]{\label{fig:3}
One-loop diagram corresponding to $\protect\rho^+
\rightarrow \protect\pi^{+} \protect\gamma$ decay.}
\end{figure}

The diagram contributing to the $\rho ^{+}(Q)\rightarrow \pi ^{+}(p)+\gamma
(q)$ decay on the one-loop level is shown in Fig.~\ref{fig:3}. The matrix
element for this process, like for any anomalous decay, has the structure
\cite{Pes95} (Ch.19.3): $\mathcal{M}=\varepsilon ^{\mu \nu \lambda \sigma
}\epsilon _{\mu }^{\ast }(\gamma )\epsilon _{\nu }(\rho )Q_{\lambda
}q_{\sigma }/(Q\cdot q)\,T,$ where $\varepsilon ^{\mu \nu \lambda \sigma }$
is the Levi-Civita antisymmetric tensor, and $T$ depends on the coupling
constants and masses of the particles. The one-loop integral corresponding
to Fig.~\ref{fig:3} converges, and using standard methods we obtain (see
Appendix~\ref{App:B})
\begin{equation}
T=(4\pi )^{-2}2eg_{\rho }g_{\pi }m_{N}\sqrt{X_{\pi }}\int_{0}^{1} \log \left[
\frac{m_{N}^{2}-m_{\rho }^{2}x(1-x)}{m_{N}^{2}-m_{\pi }^{2}x(1-x)}\right]
\frac{\mbox{\rm d}x}{1-x}.  \label{Eq.70}
\end{equation}
The calculated decay width is 55 KeV, while the PDG \cite{PDG} quotes the
value 68$\pm7$ KeV. In view of the simple mechanism assumed for this decay
we consider the agreement between the calculation and experiment
satisfactory
\footnote{The $\rho ^{0}\rightarrow \pi ^{0}\gamma $ decay width of 121$\pm31$ KeV
\cite{PDG} is not described in the present isospin symmetrical model.}.
It is worthwhile to mention that the factor $g_{\pi }g_{\rho }\sqrt{X_{\pi }},$
which defines the magnitude of the matrix element, is considerably smaller
than one would get using the conventional values for the couplings. For
example, with the typical values $g_{\pi }=13.0,$ $g_{\rho }=6.04$ (and
$X_{\pi }=1$) the width would increase to 160 KeV.

All calculated decay widths are collected in Table~\ref{tab:widths}, where
they are compared with experimental values~\cite{PDG}. We also included in
Table~\ref{tab:widths} the results~\cite{Ko94} obtained in a version of GLSM
with massive $\rho $ and $a_{1}$ mesons, where several additional terms were
introduced. In particular, the $a_{1}^{+}\rightarrow \pi ^{+}\gamma $ decay
in \cite{Ko94} appears on the tree level due to the introduction of
dimension-6 operators in the Lagrangian. Some results in the non-linear
realization of the chiral symmetry (hidden symmetry approach) from ref.~\cite
{Mei88}(Ch.3) are shown in the 7th column.

%%%%%%%%%%%%%%%%%%%%%%%%%%%%%% TABLE 2 %%%%%%%%%%%%%%%%%%%%%%%%%

\begin{table}[tbp]
\caption{\label{tab:widths}
Widths of strong and electromagnetic decay of mesons. Calculations
are performed with two masses of the $\sigma$ meson:
$m_{\protect\sigma}^{(a)}=770$ MeV and $m_{\protect\sigma}^{(b)}=1$ GeV.
Experimental values are from \protect\cite{PDG}. }
%%%\begin{ruledtabular}
\par
\begin{center}
\begin{tabular}{|l|c|c|c|c|c|c|c|}
\hline
Meson decay & \multicolumn{2}{c|}{$m_{a}=1.230$ GeV} & \multicolumn{2}{c|}{$%
m_{a}=1.089$ GeV} & Ref.~\cite{Ko94} & Ref.~\cite{Mei88} & Experiment \\
\hline
& \hspace*{.5cm}$m_{\sigma}^{(a)}$ & $m_{\sigma}^{(b)}$ & \hspace*{.5cm}$m_{\sigma}^{(a)}$ &
$m_{\sigma}^{(b)}$ &  &  &  \\ \hline
$a_{1}\rightarrow \pi \rho $ [MeV] & 272 & 272 & 163 & 163 & 483 & 360 & 150
to 361 \\
$D/S$ \ ratio [\%] & -4.6 & -4.6 & -2.3 & -2.3 & 7.8 &  & -10.7$\pm 1.6$ \\
$a_{1}\rightarrow \pi \sigma $ [MeV] & 46 & 4.7 & 21 & - &  &  & 32 to 147
\\
$\Gamma (a_{1}\rightarrow \pi \sigma )/\Gamma _{tot}$ [\%] & $\approx $ 14 &
$\approx$ 1.7 & $\approx $ 11 & - &  &  & 18.76$\pm 4.29\pm $1.48 \\
$\sigma \rightarrow \pi \pi $ [MeV] & 149 & 346 & 325 & 753 & 373 &  &  \\
$\rho ^{0}\rightarrow \pi ^{+}\pi ^{-}\gamma $ [MeV]: &  &  &  &  &  &  &
\\
$\;\;\; (k_{\min}=50$ MeV) & 1.73 & 1.73 & 1.73 & 1.73 &  &  & 1.487$
\pm0.240 $ \\
$\;\;\; (k_{\min}=60$ MeV) & 1.49 & 1.49 & 1.49 & 1.49 &  &  &  \\
$a_{1}^{+}\rightarrow \pi ^{+}\gamma $ [KeV] & 412 & 411 & 192 & 180 & 670 &
300 & 640$\pm 246$ \\
$\rho ^{+}\rightarrow \pi ^{+}\gamma $ [KeV] & 55 & 55 & 81 & 81 &  & 80 & 68
$\pm 7$ \\ \hline
\end{tabular}
%%%\end{ruledtabular}
\end{center}
\par
\end{table}

%%%%%%%%%%%%%%%%%%%%%%%%%%%%%%%%%%%%%%%%%%%%%%%%%%%%%%%%%%%%%%%%%%%%%
%%%%%%%%%%%%%%%%%%%%%%%%%%%%%%%%%%%%%%%%%%%%%%%%%%%%%%%%%%%%%%%%%%%%%

\section{\label{sec:pi-pi}$\protect\pi -\protect\pi $ scattering at low
energies}

Pion-pion scattering is a process where one can test the strong-interaction
Lagrangian. Let us first analyze the terms in the Lagrangian in
Eqs.(\ref{Eq.c42}) and (\ref{Eq.c43}) which are relevant for the $\pi -\pi $ scattering on the
tree level:
\begin{eqnarray}
\Delta \mathcal{L} &=&-\lambda X_{\pi }
\mbox{\boldmath $\pi$\unboldmath}^{2}(v\sigma +\frac{1}{4}X_{\pi }
\mbox{\boldmath $\pi$\unboldmath}^{2})-
\frac{1}{2}g_{\rho }(1+X_{\pi })\mbox{\boldmath $\rho$\unboldmath}_{\mu }
(\mbox{\boldmath $\pi$\unboldmath}\times \partial ^{\mu }\mbox{\boldmath
$\pi$\unboldmath})  \nonumber \\
&&-\frac{\lambda _{H}}{8}(1-X_{\pi })
\mbox{\boldmath $\pi $\unboldmath}^{2}[u\eta +\frac{1}{4}(1-X_{\pi })
\mbox{\boldmath $\pi $\unboldmath}^{2}].
\label{Eq.71}
\end{eqnarray}
This Lagrangian has several unusual features. Firstly, the $\sigma \pi \pi $
interaction and the related $\pi ^{4}$ interaction, coming from the first
term in Eq.(\ref{Eq.71}), are suppressed by the factors $X_{\pi }$ and \
$X_{\pi }^{2}$ respectively. This will lead to a suppression of the
corresponding amplitude by the factor $X_{\pi }^{2}\approx 0.15.$ Secondly,
the $\rho $-meson exchange is determined by the coupling $g_{\rho \pi \pi
}=g_{\rho }(1+X_{\pi })/2$ which is fixed from the $\rho \rightarrow \pi \pi
$ decay. Thirdly, the last term, containing $\eta \pi \pi $ and $\pi ^{4}$
interactions, has a coupling $\lambda _{H}$ which rises with the Higgs mass
$m_{H}$ (see Eq.(\ref{Eq.22})). At first sight this leads to a divergence of
the tree-level amplitude in the limit $m_{H}\rightarrow \infty$, and it
seems unlikely that the Lagrangian (\ref{Eq.71}) can give reasonable
predictions for $\pi -\pi $ scattering. In this section we will study this
issue by calculating the low-energy scattering parameters for the $S-$ and
$P-$waves.

The formalism of $\pi -\pi $ scattering has been considered in many
references (see, \textit{e.g.} \cite{Ber91,Ko94}), and we briefly
recall the basic relations. The scattering amplitude for the reaction \
$\pi ^{i}+\pi ^{j}\rightarrow \pi ^{k}+\pi ^{l}$ \ can be written as
\begin{equation}
\mathcal{M}_{ij,kl}=M(s,t,u)\delta ^{ij}\delta ^{kl}+M(t,s,u)\delta
^{ik}\delta ^{jl}+M(u,t,s)\delta ^{il}\delta ^{jk},  \label{Eq.72}
\end{equation}
where $i,j,k,l$ label the charge states of the pions, and $s,t,u$ are the
conventional Mandelstam variables. One also defines the amplitudes
$M^{(I)}$ \ with total isospin $I=0,1,2:$
\begin{eqnarray}
M^{(0)} &=&3M(s,t,u)+M^{(2)},\,\ \ \ \ \ \ \ \ \ \ \ \ \ \ \ \
M^{(1)}=M(t,s,u)-M(u,t,s),  \nonumber \\
M^{(2)} &=&M(t,s,u)+M(u,t,s).
 \label{Eq.73}
\end{eqnarray}
The amplitude in a channel with fixed isospin is expanded in partial waves,
\begin{equation}
M^{(I)}=32\pi \sum_{l=0}^{\infty }(2l+1)P_{l}(\cos \theta )t_{l}^{(I)}(s),
\label{Eq.74}
\end{equation}
where $P_{l}(\cos \theta )$ is the Legendre polynomial \ and $\theta $ is
the scattering angle. The partial-wave amplitude $t_{l}^{(I)}(s)$ can be
approximated at small center-of-mass (CM) momentum $\mathbf{|q|}=(s/4-m_{\pi
}^{2})^{1/2}$ \ as follows,
\begin{equation}
\mbox{\rm Re}\ t_{l}^{(I)}(s)\approx \frac{\mathbf{q}^{2l}}{m_{\pi }^{2l}}
[a_{l}^{(I)}+b_{l}^{(I)}\frac{\mathbf{q}^{2}}{m_{\pi }^{2}}+\mathcal{O}(\frac{%
\mathbf{q}^{4}}{m_{\pi }^{4}})].  \label{Eq.75}
\end{equation}
In order to find the scattering lengths $a_{l}^{(I)}$ and effective ranges$\
b_{l}^{(I)}$ one has to expand the amplitudes in Eq.(\ref{Eq.73}) around
$\mathbf{q}^{2}=0$, using the definition of the invariants in the CM frame
\begin{equation}
s=4(m_{\pi }^{2}+\mathbf{q}^{2}),\,\ \ \ \ \ \ \ \ \ \ \ \ \ \ \ \ \ \
t=-2\mathbf{q}^{2}(1-\cos \theta ),\,\ \ \ \ \ \ \ \ \ \ \ \ \ \ \ \
u=-2\mathbf{q}^{2}(1+\cos \theta ).  \label{Eq.76}
\end{equation}

We now show that the Higgs part of the Lagrangian (last term in Eq.(\ref
{Eq.71})) gives a finite contribution to $M(s,t,u)$. The $\eta $ exchange
and $\pi ^{4}$ interaction lead to the amplitude
\begin{equation}
M_{\eta }(s,t,u)=\left( -\frac{\lambda _{H}}{4}+\frac{1}{16}\frac{\lambda
_{H}^{2}u^{2}}{m_{H}^{2}-s}\right) (1-X_{\pi })^{2}=\frac{g_{\rho
}^{2}(1-X_{\pi })^{2}}{4m_{\rho }^{2}}\frac{s}{(1-s/m_{H}^{2})},
\label{Eq.77}
\end{equation}
where we made use of $\lambda _{H}u^{2}=4m_{H}^{2}$, which follows from
Eqs.(\ref{Eq.22}). It is clear that the couplings in the $\eta $ exchange and the
contact $\pi ^{4}$ diagram are tuned in such a way that the sum remains
finite at large $m_{H}.$

It is straightforward to obtain from Eq.(\ref{Eq.71}) the amplitudes
corresponding to $\sigma $ exchange, with the associated $\pi ^{4}$ term,
and $\rho $ exchange,
\begin{equation}
M_{\sigma }(s,t,u)+M_{\rho }(s,t,u)=-2\lambda X_{\pi }^{2}\frac{m_{\pi
}^{2}-s}{m_{\sigma }^{2}-s}+g_{\rho \pi \pi }^{2}\left( \frac{s-u}{m_{\rho
}^{2}-t}+\frac{s-t}{m_{\rho }^{2}-u}\right) .  \label{Eq.78}
\end{equation}
The total amplitude is $\ M(s,t,u)=$\ $M_{\eta }(s,t,u)+M_{\sigma
}(s,t,u)+M_{\rho }(s,t,u).$ The other amplitudes in Eq.(\ref{Eq.72}) are
obtained by interchanging $t\longleftrightarrow s,$ or $u\longleftrightarrow
s.$ As expected, the $\sigma $ contribution in Eq.(\ref{Eq.78})
is multiplied by $X_{\pi }^{2}$ which strongly reduces the effect of
the $\sigma $ meson in $\pi -\pi $ scattering.
The low-energy parameters can be found by expanding Eq.(\ref
{Eq.77}) and (\ref{Eq.78}) in powers of $\mathbf{q}^{2}$, using the
definitions (\ref{Eq.74}),(\ref{Eq.76}), and comparing the results
with Eq.(\ref{Eq.75}). The calculated coefficients
$a_{l}^{(I)}$ and $b_{l}^{(I)}$\ are presented in Table~\ref{tab:pi-pi}.

\begin{table}[tbp]
\caption{\label{tab:pi-pi}
Low-energy parameters for $\protect\pi -\protect\pi$ scattering.
The $\protect\sigma$-meson mass is 770 MeV, and the $a_1$-meson mass is 1.23
GeV. }
%%%\begin{ruledtabular}
\par
\begin{center}
\begin{tabular}{|l|l|l|l|l|l|}
\hline
Model & $a_{0}^{(0)}$ & $b_{0}^{(0)}$ & $a_{0}^{(2)}$ & $b_{0}^{(2)}$ & $%
a_{1}^{(1)}$ \\ \hline
Present model: $\sigma$ + $\rho$ + $\eta$ & 0.238 & 0.319 & -0.100 & -0.155
& 0.061 \\
-- $\;\;\;\;\;\;\;\;\;\;\;\;\;$ only $\sigma$ + $\rho $ & 0.210 & 0.301 &
-0.100 & -0.146 & 0.057 \\
-- $\;\;\;\;\;\;\;\;\;\;\;\;\;$ only $\rho $ & 0.191 & 0.273 & -0.095 &
-0.137 & 0.055 \\
-- with form factor for $\rho $ exchange & 0.160 & 0.200 & -0.061 & -0.097 &
0.040 \\
$\;\;\;\;\;\;\;\;\;\;\;\;\; $($\Lambda =1.6$ GeV) &  &  &  &  &  \\
Soft-pion amplitude \cite{Wei66} & 0.159 & 0.182 & -0.045 & -0.091 & 0.030
\\
ChPT (only pions) \ \cite{Gae83} & 0.20 & 0.24 & -0.042 & -0.075 & 0.037 \\
ChPT (pions and resonances) \cite{Ber91} & 0.21 &  & -0.043 &  & 0.038 \\
\hline
Experiment \ \cite{Nag79,Rig91} & 0.26$\pm 0.05$ & 0.25$\pm 0.03$ & -0.028$%
\pm 0.012$ & -0.082$\pm 0.008$ & 0.038$\pm 0.002$ \\
Experiment \ \cite{Sev92} & 0.20$\pm 0.01$ &  & -0.037$\pm 0.004$ &  &  \\
\hline
\end{tabular}
%%%\end{ruledtabular}
\end{center}
\end{table}

It is seen from Table~\ref{tab:pi-pi} that the scattering length
$a_{0}^{(0)} $ is described fairly well. The other parameters, however, are
overpredicted by a factor of 1.5$\div$2. The $\rho $ meson gives the
dominant contribution (compare the 2nd and 4th rows), while the contribution
of the Higgs meson is small (see the 2nd and 3d rows). In the 6th row we
show parameters obtained with the soft-pion amplitude \
$M_{SPA}(s,t,u)=(s-m_{\pi }^{2})\,/f_{\pi }^{2}$ \ from ref.~\cite{Wei66},
which is based on PCAC and current commutation relations. Our amplitude will
reduce to $M_{SPA}(s,t,u)$ if $\ g_{\rho }=0$ and \ $m_{\sigma }^{2}\gg
m_{\pi }^{2}.$ \ We also show results of the ChPT calculations:
\cite{Gae83}, where only pions are included and \cite{Ber91},
where resonances are added.

Comparison with the experiment indicates that the effect of the $\rho $
meson is overemphasized in the present model. In this connection we compare
the $\rho $-exchange amplitude in Eq.(\ref{Eq.78}) with the corresponding
amplitude by Bernard {\em et al.} [Eq.(3.14) in ref.~\cite{Ber91}]. One of the
differences is that
in \cite{Ber91} the propagator of the vector meson (to be exact, the part
contributing for on-shell pions) is modified according to
\begin{equation}
\frac{1}{m_{\rho }^{2}-t} \longrightarrow
\frac{t}{m_{\rho }^{2}(m_{\rho }^{2}-t)}
= \frac{1}{m_{\rho }^{2}-t}-\frac{1}{m_{\rho }^{2}},
\label{Eq.79}
\end{equation}
which coincides with Weinberg's suggestion \cite{Wein68} based on
chiral-symmetry arguments. Such a modification strongly reduces the effect
of the $\rho $ meson at low energies. In the present model however there is
no compensating term of the form~\cite{Wein68}
\ $ \sim(\mbox{\boldmath$\pi$\unboldmath}\times \partial _{\mu }
\mbox{\boldmath $\pi$\unboldmath})^{2} $\ which would lead to Eq.(\ref
{Eq.79}). Using the $\pi \pi \rho $ interaction with more derivatives, as is
advocated in ref.~\cite{Ber91}, would also be inconsistent with the
Lagrangian (\ref{Eq.c42}). Besides, the calculation shows that the effect of
the $\rho $ meson cannot be eliminated completely as would follow from
Eq.(\ref{Eq.79}), since the $\sigma $ contribution in itself is by far too
small. One of the mechanisms which can partly reduce the $\rho $
contribution is the dependence of the $\rho \pi \pi $ vertex on the
invariant mass of the $\rho .$ One may assume that $\ g_{\rho \pi \pi
}(t)=g_{\rho \pi \pi }F_{\rho \pi \pi }(t)$ \ with a form factor normalized
to unity at $t=m_{\rho }^{2}.$ The typical form often used in
phenomenological models is \ $F_{\rho \pi \pi }(t)=(\Lambda ^{2}-m_{\rho
}^{2})/(\Lambda ^{2}-t),$ where $\Lambda $ is a cut-off parameter. Choosing
the value $\Lambda =1.6$ GeV \cite{Loh90} we obtain a considerable reduction
of the low-energy parameters (see Table~\ref{tab:pi-pi}, the 5th row). Of
course this is only one plausible argument which may explain the
unsatisfactory description of low-energy $\pi -\pi $ scattering on the tree
level. Besides the form factor, other higher-order corrections to the
amplitude would have to be consistently included. This issue will be studied
elsewhere.

%%%%%%%%%%%%%%%%%%%%%%%%%%%%%%%%%%%%%%%%%%%%%%%%%%%%%%%%%%
%%%%%%%%%%%%%%%%%%%%%%%%%%%%%%%%%%%%%%%%%%%%%%%%%%%%%%%%%%

\section{\label{sec:Discussion}Discussion and conclusions}

In the previous sections we have presented the results of calculations in
the framework of a new representation of QHD-III \cite{Pre99}. An advantage
of this new form, compared to the original formulation \cite{Ser92}, is
that only simple vertices with at most one derivative appear in the
Lagrangian. As a result the calculations are simpler, and more importantly,
the strong-interaction vertices describing the decay of the mesons do not
vanish at any values of the meson invariant mass. A comparison with
experiment (Table~\ref{tab:widths}) shows that most of the decay widths are
reasonably described. This was not anticipated in view of the fact that
practically no free parameters were used in the calculations. In fact, only
the coupling $g_{\rho }$ was fixed from the $\rho \rightarrow \pi \pi $
decay. The masses of the nucleon, pion, rho and $a_{1}$ mesons were taken
from the latest PDG review \cite{PDG}.

The mass of the $\sigma $ meson was chosen equal to the mass of the
$\rho ,$ \textit{i.e.}
$m_\sigma =m_\rho =770$ MeV, in line with ref.~\cite{Wei90}
where the $\sigma $  is supposed to be degenerate with the $\rho$.
With this mass the calculated width for the $a_1 \rightarrow \pi \sigma$ decay,
46 MeV, is surprisingly close to the value predicted in \cite{Wei90}:
$\Gamma (a_{1}\rightarrow \pi \sigma )=\Gamma (\rho \rightarrow \pi \pi
)\,/\sqrt{8}\approx 50$ MeV.
In some calculations we use the value $m_{\sigma }=$ 1 GeV.
As a result the widths of the $a_1 \rightarrow \pi \sigma$ and
$\sigma \rightarrow \pi \pi$ decays change considerably due to a change of the
available phase space.
The other observables are not sensitive to the $\sigma $ mass.
Calculations were also
performed with the value $m_{a}=\sqrt{2}m_{\rho }=1.089$ GeV, which is based
on different arguments \cite{Gil68,Wei90}. However, agreement with
experiment is better with the PDG value $m_{a}=1.23$ GeV. As a curious
observation we mention that this value approximately obeys the
relation $m_{a}\approx (1+\sqrt{2})^{1/2}m_{\rho }$ within a 3\% accuracy. This
relation also implies that $g_{\rho }/g_{\rho \pi \pi
}=2m_{a}^{2}/(m_{a}^{2}+m_{\rho }^{2}),$ \textit{i.e.} the ratio of the
$\rho $ couplings to the nucleon and the pion, is very close to $\sqrt{2}.$

The EM decay $\rho ^{0}\rightarrow \pi ^{+}\pi ^{-}\gamma $ is described by
the two-step tree-level amplitude supplemented by a contact diagram. The
latter comes from the \ $\gamma \pi \pi \rho $ vertex in the EM current (\ref
{Eq.c39}) and guarantees the EM gauge invariance. The decay is mainly
determined by the $\rho \pi \pi $ coupling and is not
sensitive to other ingredients of the model.
The decay $a_{1}^{+}\rightarrow
\pi ^{+}\gamma $ is a more informative process. As there is no direct $a\pi
\gamma $\ coupling the process is described by many one-loop diagrams which
strongly interfere. The total amplitude of $a_{1}^{+}\rightarrow \pi
^{+}\gamma $ comes out finite despite the divergence of the separate
diagrams. Since the Lagrangian was obtained in the unitary gauge, the
propagator of the vector mesons includes the longitudinal component $k^{\mu
}k^{\nu }/m^{2}.$ The latter gives rise to a quadratic divergence, $\sim
\Gamma (1-d/2)$, but due to the EM gauge invariance these divergences from
different diagrams cancel. Furthermore, the logarithmic divergences, $\sim
\Gamma (2-d/2)$, cancel as well.

The Higgs mesons, $\eta $ and $\zeta$, play the role of auxiliary particles
in this model. According to ref.~\cite{Ser92} these mesons serve as
regulators and should have minimal effect on the low-energy predictions.
They do not appear in the initial and final states, but may contribute as
intermediate particles.
For instance, the $\eta $ and $\zeta $ appear in one-loop diagrams
for the $a_{1}^{+}\rightarrow \pi ^{+}\gamma $ decay.
In the diagrams with intermediate $\pi^+ \eta$ state, the
$\pi \pi \eta $ coupling in Eq.(\ref{Eq.c43}) increases proportionally
to $m^2_{H}.$ However, due to the presence of the $\eta$ propagator this
contribution  stays finite.
The situation is different in the diagrams with an intermediate $a^+ \eta$
and $\rho^+ \zeta$,
where the vertices $a a \eta$ and $a \rho \zeta$ are independent of $m_H$.
Nevertheless the amplitudes do not vanish in the limit $m_H \rightarrow \infty$,
as one would naively expect by taking the limit of the propagator in the loop
integrand. This is because the longitudinal component of the vector meson propagator
leads to divergent integrals; the correct procedure is to take the limit
$m_H \rightarrow \infty $ after the loop integration, and leads to a nonzero
contribution.
At $m_H \rightarrow \infty$
the total amplitude corresponding to the above processes involving Higgs mesons
takes a form equivalent  to an effective $a \pi \gamma$ Lagrangian.
Numerically its contribution to the
$a_{1}^{+}\rightarrow \pi ^{+}\gamma$ decay turns out to be relatively small.

In processes where $\eta $ and $\zeta$ appear on the tree level
the amplitude may diverge at $m_{H}\rightarrow \infty$, if
there are more $\pi \pi \eta $ vertices than the $\eta$ propagators.
Using the $\pi -\pi $ scattering as an example we
demonstrated that this is not the case. In this reaction
there is the $\eta$-exchange diagram which at first sight behaves as $m_{H}^2$.
However there is a compensating $\pi ^{4}$ vertex in the
Lagrangian (\ref{Eq.c43}). The couplings in the $\eta $ exchange and $\pi
^{4}$ contact diagrams are tuned in such a way that the sum remains finite.
This mechanism of cancellation is similar to that in the linear sigma model,
in which the amplitude given by the $\sigma $ exchange and the $\pi ^{4}$
vertex does not diverge in the limit $m_{\sigma }\rightarrow \infty $, but
rather takes the value dictated by the soft-pion amplitude \cite{Wei66}. The
contribution of the $\eta $ exchange and the associated $\pi ^{4}$ term at
low energies is small.

The pion interaction with hadrons in the model is scaled down by the factor $%
\sqrt{X_{\pi }}=m_{\rho }/m_{a}$ (with the exception of the $\pi \pi \rho $,
$\pi \pi \rho \rho $ and $\pi \pi aa$ vertices). This influences the matrix
elements of the meson decays, and in most cases improves agreement with
experiment. The factor $m_{\rho }/m_{a}$\ also leads to a reduction of
the $\sigma \rightarrow \pi \pi $ decay width. For example, compared to the
linear sigma model the width is reduced by almost an order of magnitude.
Therefore the width comes out smaller than the $\sigma $ mass [of the order
of 150 (350) MeV for $m_\sigma=$0.77 (1.0) GeV]
which may be helpful in identifying the $\sigma $ with the
scalar-isoscalar state around 1 GeV \cite{PDG}. At the same time the effect
of the $\sigma $ in $\pi -\pi $ scattering is also diminished. Our
calculation shows that the $\sigma $ contribution to the low-energy
parameters becomes very small, and the dominant contribution comes from the
$\rho $ exchange. The agreement between the calculation and experiment is not
very impressive compared to, for example, ChPT calculations. The scattering
length in the $I=0,\ l=0$ channel is described quite well, but the other
calculated parameters overestimate the experiment. This deficiency may be
related to the tree-level approximation. We included a form factor in the $\pi
\pi \rho $ vertex, similarly to what is often done in phenomenological
models. This is one of the effects which contribute beyond the tree level.
In this way the $\rho $ contribution is reduced and the agreement is
improved, but other higher-order corrections need to be consistently taken
into account before definite conclusions can be drawn.

In this paper we focused on the meson properties and left out the nucleon
sector. Inclusion of the latter can also be an important test of the model,
especially because the nucleon-pion vertex is reduced. At this point we
would like to mention that we did not include an additional baryon, the
cascade $\Xi =(\Xi ^{0},\Xi ^{-})$ with the hypercharge $Y_{\Xi }=-1$. This
isodoublet was added in \cite{Ser92} to cure the problem with the
chiral anomaly and render the theory renormalizable. The chiral anomaly in
QHD-III will show up when the isoscalar $\omega $ meson couples to the
baryon loop with other two isovector vertices, one of which contains $\gamma
_{5}$. Such processes were not considered here. The anomaly may however be
important in the EM decay\ $\rho ^{+}\rightarrow \pi ^{+}\gamma $ considered
in Sect.~\ref{subsubsec:rho-pi-gam}. If we added the loop with the cascade
then the calculated width would change, though we did not include this effect.
It seems logical first to extend the $SU(2)_{R}\times SU(2)_{L}$ model to the
strange sector, as for example, was done in ref.~\cite{Gas69} for GLSM. The
extension to $SU(3)_{R}\times SU(3)_{L}$ would allow one to include along
with $\Xi $ the other strange hadrons, like $\Lambda ,\Sigma $ and $K.$

\bigskip
In conclusion, we applied chiral quantum hadrodynamics
(QHD-III) \cite{Ser92} in the calculation of some properties of the mesons.
First we included electromagnetic interaction by extending the symmetry to
the local $U(1)\times SU(2)_{R}\times SU(2)_{L}$ group. This allowed us to
obtain consistently the minimal and nonminimal contributions to the
electromagnetic interaction in an arbitrary gauge. After an appropriate
diagonalization and fixing the gauge the Lagrangian is obtained in terms of
the physical pion field. We calculated the strong and EM decays
of the vector and axial-vector mesons, $a_{1}\rightarrow \pi \rho $, $a_{1}\rightarrow\pi \sigma$, $\rho
^{0}\rightarrow \pi ^{+}\pi ^{-}\gamma $, $a_{1}^{+}\rightarrow \pi
^{+}\gamma $, $\rho ^{+}\rightarrow \pi ^{+}\gamma $, and
addressed the issue of the width of the $\sigma $ meson.
For the $a_{1}^+ \rightarrow
\pi^{+} \gamma$ decay some loop diagrams are not yet included,
and a more complete analysis is reserved for future work.
Most of the calculated decay widths are in reasonable agreement
with experiment \cite{PDG}. The only free parameter used in calculations is
the mass of the $\sigma $ meson, although most of the results
were not sensitive to $m_{\sigma }.$

We studied the effect of auxiliary Higgs bosons of QHD-III in
the $a_{1}^{+}\rightarrow \pi ^{+}\gamma $ decay and $\pi -\pi $ scattering.
The contribution of these particles to the amplitude, both on the tree level
and in the one-loop diagrams, turns out to be finite and small in the
limit $m_{H}\rightarrow \infty $. This goes in line with the viewpoint
of ref.~\cite{Ser92} on the role of $\eta $ and $\zeta $\ mesons in
low-energy hadron physics.

Our exploratory study shows that QHD-III in the representation of ref.\cite{Pre99} can
describe some features of meson phenomenology in the non-strange sector.
There are many interesting issues which can be further addressed, such as an
extension of calculations to the baryon sector ($\pi -N$ scattering and
nucleon form factors), a clarification of the role of the cascade $\Xi $ and
inclusion of the other strange hadrons. Finally, applications to the many-body sector, i.e. nuclear matter and finite nuclei, may be considered. As was shown in ref.~\cite{Fur93} (see also \cite{Ser97}, sect. 3A), the linear sigma model, when applied to finite nuclei on the mean-field level, has some deficiencies. The present model is much richer and differs in several respects, such as the presence of additional particles and vertices, and a considerable suppression of the pseudo-scalar pion-nucleon coupling. Detailed calculation will help to
assess the applicability of the model to the many-body sector.

%%%%%%%%%%%%%%%%%%%%%%%%%%%%%%%%%%%%%%%%%%%%%%%%%%%%%%%%

\begin{acknowledgments}

We would like to thank Gary Pr\'{e}zeau for clarifying
communication regarding ref.~\cite{Pre99}, and Olaf Scholten for useful discussions.
This work was supported by the Fund for Scientific Research-Flanders
(FWO-Vlaanderen) and the Research Board of Ghent University.
\end{acknowledgments}

%%%%%%%%%%%%%%%%%%%%%%%%%%%%%%%%%%%%%%%%%%%%%%%%%%%%%%%%%%%%

\appendix

\section{\label{App:A} Derivation of the Lagrangian in the Higgs sector in
an arbitrary gauge}

The most complicated part of the Higgs Lagrangian are the terms with the
covariant derivatives. For the right field, for example, such a term can be
written in the form
\begin{equation}
(D_{\mu }\Phi _{R})^{\dagger }(D^{\mu }\Phi _{R})=\frac{1}{4}\left(
\begin{array}{cc}
0 & 1
\end{array}
\right) [Y_{1}\cdot Y_{1}+Y_{2}\cdot Y_{2}+Y_{3}\cdot \tilde{Y}%
_{3}-Y_{1}\cdot (Y_{3}+\tilde{Y}_{3})+i(Y_{2}\cdot \tilde{Y}_{3}-Y_{3}\cdot
Y_{2})]\left(
 \begin{array}{c}
0 \\
1
\end{array}
\right) \,,  \label{Eq.A1}
\end{equation}
where the operators $Y_{1},Y_{2},Y_{3}$ and \ $\tilde{Y}_{3}$\ are defined
as
\begin{eqnarray}
Y_{1}^{\mu } &=&(\partial ^{\mu }\eta +\partial ^{\mu }\zeta
),\;\;\;\;\;\;\;\;Y_{2}^{\mu }=\mbox{\boldmath $\tau$\unboldmath}(\partial
^{\mu }\mathbf{H+\partial }^{\mu }\mathbf{Z})+\frac{1}{2}(u+\eta +\zeta
)Y_{4}^{\mu },\;\;\;\;\;  \nonumber \\
Y_{3}^{\mu } &=&\frac{1}{2}\mbox{\boldmath $\tau$\unboldmath}(\mathbf{H+Z}%
)Y_{4}^{\mu },\;\;\;\;\;\;\;\;\tilde{Y}_{3}^{\mu }=\frac{1}{2}Y_{4}^{\mu }%
\mbox{\boldmath$\tau$\unboldmath}(\mathbf{H+Z}),  \nonumber \\
Y_{4}^{\mu } &=&g^{\prime}_{\rho }\mbox{\boldmath $\tau$\unboldmath}(%
\mbox{\boldmath
$\rho$\unboldmath}^{\prime \mu }+\mathbf{a}^{\mu })+e^{\prime}A^{\prime \mu }\approx
g_{\rho }\mbox{\boldmath $\tau$\unboldmath}({%
\mbox{\boldmath
$\rho$\unboldmath}}^{\mu }+\mathbf{a}^{\mu })+e{A}^{\mu }(1+\tau_{3}),
\label{Eq.A3}
\end{eqnarray}
and the notation $X\cdot Y=X_{\mu }Y^{\mu }$ is used. The similar term for the
field $\Phi _{L}$ can be obtained from the above formulas by changing \ $%
\zeta \rightarrow -\zeta ,$ $\mathbf{Z}\rightarrow -\mathbf{Z}$ and $\mathbf{%
a}_{\mu }\rightarrow -\mathbf{a}_{\mu }.$ The potential $V_{H}$ in Eq.(\ref
{Eq.3}) is chosen such that $\lambda _{H}>0$ and $\mu _{H}^{2}>0,$ thus
allowing for SSB of the global gauge invariance \cite{Ser92}.
From the condition that the term linear in $\eta $ must be absent,
the VEV $u$ can be found. The fields $\eta $ and $\zeta $ acquire a mass $m_{H}$,
whereas $\mathbf{H}$ and $\mathbf{Z}$ remain massless, indicating that these are the
Goldstone bosons.

Calculation of the matrix elements in Eq.(\ref{Eq.A1}) leads to the EM
current in Eq.(\ref{Eq.16}) and (\ref{Eq.17}). The strong Lagrangian
consists of the free Lagrangians in Eqs.(\ref{Eq.20}-\ref{Eq.21}), the mixing
term in Eq.(\ref{Eq.19}), and the interaction Lagrangian
\begin{eqnarray}
\mathcal{L}_{H}^{int} &=&\frac{1}{8}g_{\rho }^{2}
\{(\mbox{\boldmath
$\rho$\unboldmath}_{\mu }^{2}+\mathbf{a}_{\mu }^{2})(\eta ^{2}+\zeta ^{2}+%
\mathbf{H}^{2}+\mathbf{Z}^{2}+2u\eta )
+4 \mbox{\boldmath$\rho$\unboldmath}_{\mu }
\mathbf{a}^{\mu}[(u+\eta )\zeta +\mathbf{HZ}] \}
 \nonumber \\&&
+\frac{1}{2}g_{\rho }[\mbox{\boldmath $\rho$\unboldmath}_{\mu }(\eta
\partial ^{\mu }\mathbf{H}-\mathbf{H}\partial ^{\mu }\eta +\zeta \partial
^{\mu }\mathbf{Z}-\mathbf{Z}\partial ^{\mu }\zeta -\mathbf{H}\times \partial
^{\mu }\mathbf{H}-\mathbf{Z}\times \partial ^{\mu }\mathbf{Z})+\mathbf{a}
_{\mu }(\eta \partial ^{\mu }\mathbf{Z}-\mathbf{Z}\partial ^{\mu }\eta
\nonumber \\
&&+\zeta \partial ^{\mu }\mathbf{H}-\mathbf{H}\partial ^{\mu }\zeta -\mathbf{
H}\times \partial ^{\mu }\mathbf{Z}-\mathbf{Z}\times \partial ^{\mu }\mathbf{
H})]-v_{H}(\eta ,\zeta ,\mathbf{H,Z})\,,
\label{Eq.A-Higgs-strong}
\end{eqnarray}
where the remaining piece of the potential is
\begin{eqnarray}
v_{H}(\eta ,\zeta ,\mathbf{H,Z}) &=&\frac{\lambda _{H}}{32}[\eta ^{4}+\zeta
^{4}+4u\eta ^{3}+6\eta \zeta ^{2}(2u+\eta )+\mathbf{H}^{4}+\mathbf{Z}^{4}
+2\mathbf{H}^{2}\mathbf{Z}^{2}  \nonumber \\
&&+4(\mathbf{HZ})^{2}+2(\mathbf{H}^{2}+\mathbf{Z}^{2})(\eta ^{2}+\zeta
^{2}+2u\eta )+8(u+\eta )\zeta \mathbf{HZ}]\,.
  \label{Eq.A-V-Higgs}
\end{eqnarray}
Eqs.(\ref{Eq.A-Higgs-strong}-\ref{Eq.A-V-Higgs}) will reduce to the
Lagrangian of QHD-III \cite{Ser92} if we take $\mathbf{H}=\mathbf{Z}=0.$

The interaction Lagrangian of GLSM from Sect.~\ref{subsec:Sigma} follows
from Eqs.(\ref{Eq.23}) - (\ref{Eq.26}) after we define the $\sigma$ field
through $\phi =v+\sigma\ $, where the VEV $v$ can be fixed from the minimum of
the potential, $\lambda v^{3}-\mu ^{2}v-c=0$. The Lagrangian has the form
\begin{eqnarray}
\mathcal{L}_{N\pi \sigma \omega }^{int} &=&-\bar{N}[g_{\pi }(\sigma +i\gamma _{5} %
\mbox{\boldmath $\tau$\unboldmath}\mbox{\boldmath $\pi$\unboldmath})
+g_{\rho}\gamma ^{\mu }\frac{\mbox{\boldmath $\tau$\unboldmath}}{2} (%
\mbox{\boldmath
$\rho$\unboldmath}_{\mu }+\gamma _{5}\mathbf{a}_{\mu })+g_{\omega }\gamma
^{\mu }\omega _{\mu }]N  \nonumber \\
&& +g_{\rho }[\mathbf{a}_{\mu }(\mbox{\boldmath $\pi$\unboldmath}\partial
^{\mu }\sigma -\sigma \partial ^{\mu }\mbox{\boldmath $\pi$\unboldmath})- %
\mbox{\boldmath $\rho$\unboldmath}_{\mu }(\mbox{\boldmath $\pi$\unboldmath}
\times \partial ^{\mu }\mbox{\boldmath $\pi$\unboldmath}) -g_{\rho } va_{\mu
} (\mbox{\boldmath $\pi$\unboldmath}\times
\mbox{\boldmath
$\rho$\unboldmath}^{\mu }-\sigma \mathbf{a}^{\mu })]  \nonumber \\
&& +\frac{1}{2}g_{\rho}^{2}[(\mathbf{a}_{\mu }
\mbox{\boldmath
$\pi$\unboldmath})^{2} +(\mbox{\boldmath
$\pi$\unboldmath}\times \mbox{\boldmath $\rho$\unboldmath}_{\mu }-\sigma
\mathbf{a}_{\mu })^{2}] -\lambda (\sigma ^{2}+%
\mbox{\boldmath
$\pi$\unboldmath}^{2})[v\sigma +\frac{1}{4}(\sigma ^{2}+%
\mbox{\boldmath
$\pi$\unboldmath}^{2})]\,.  \label{Eq.A-N-strong}
\end{eqnarray}

%%%%%%%%%%%%%%%%%%%%%%%%%%%%%%%%%%%%%%%%%%%%%%%%%%%%
%%%%%%%%%%%%%%%%%%%%%%%%%%%%%%%%%%%%%%%%%%%%%%%%%%%%
%%%%%%%%%%%%%%%%%%%%%%%%%%%%%%%%%%%%%%%%%%%%%%%%%%%%

\section{\label{App:B} Evaluation of loop integrals}

Let us consider the amplitudes for $a_{1}^{+}\rightarrow \pi ^{+}\gamma $
decay with an intermediate $(\pi ^{+}\rho ^{0})$ state, corresponding to
diagrams in Fig.~\ref{fig:2} with $j=1$. We introduce the following
notation:
\begin{equation}
D_{k}=k^{2}-m_{\rho }^{2},\,\ \ \ \ \ \ \ \ \ \ \ d_{k+Q}=(k+Q)^{2}-m_{\pi
}^{2},\,\ \ \ \ \ \ \ \ \ \ \ d_{k+p}=(k+p)^{2}-m_{\pi }^{2}  \label{Eq.B1}
\end{equation}
with $Q^{2}=m_{a}^{2},\,\ p^{2}=m_{\pi }^{2}$ and $\ q^{2}=0.$
The amplitude labeled $(a)$ in Fig.~\ref{fig:2} has an integrand proportional to
\begin{eqnarray}
M_{1a}^{\mu \nu } &\Longrightarrow& (2k+2Q-q)^{\mu }(2p+k)^{\sigma }(g_{\nu
\sigma }- \frac{k_{\nu }k_{\sigma }}{m_{\rho }^{2}}  )\frac{1}{D_{k}\ d_{k+Q}\,d_{k+p}}
\nonumber \\
&\Longrightarrow &\frac{2(k+Q)^{\mu }}{D_{k}\ d_{k+Q}}\left[ \frac{k^{\nu }}{%
m_{\rho }^{2}}-\frac{(2p+k)^{\nu }}{d_{k+p}}\right] \, .  \label{Eq.B2}
\end{eqnarray}
The terms proportional to $Q^{\nu }$ and $q^{\mu }$ are omitted hereafter because of
the relations $Q\cdot \epsilon (a)=0$ and $q\cdot \epsilon (\gamma )=0.$ The
diagram $(b)$ in Fig.~\ref{fig:2} has an integrand
\begin{equation}
M_{1b}^{\mu \nu }\Longrightarrow \frac{2}{D_{k}\ d_{k+Q}}\left( g^{\mu \nu }-
\frac{k^{\mu }k^{\nu }}{m_{\rho }^{2}}\right) .
\label{Eq.B3}
\end{equation}
For the diagram $(c)$ with photon bremsstrahlung off the $a_{1}$ meson we
obtain after some algebra,
\begin{eqnarray}
M_{1c}^{\mu \nu } &\Longrightarrow &\frac{1}{q\cdot Q\,D_{k}} \left[
\frac{Q^{\mu}(k-2q)^{\nu }+(k+2Q)^{\mu }q^{\nu }-g^{\mu \nu }q\cdot (2Q+k)}
{d_{k+p}} \right.  \nonumber \\
&&\left. -\frac{Q^{\mu }k^{\nu }+k^{\mu }q^{\nu }-g^{\mu \nu }q\cdot k}
{m_{\rho }^{2}}\right] .  \label{Eq.B4}
\end{eqnarray}
Similarly we find for the diagram $(d)$ in Fig.~\ref{fig:2} with
bremsstrahlung off the pion,
\begin{equation}
M_{1d}^{\mu \nu }\Longrightarrow \frac{Q^{\mu }k^{\nu }}{D_{k}}\left( \frac{1%
}{q\cdot Q\,\,m_{\rho }^{2}\ }-\frac{1}{q\cdot Q\,\ d_{k+Q}}
-\frac{2}{m_{\rho }^{2}\ d_{k+Q}}\right) .
\label{Eq.B5}
\end{equation}
In the chosen unitary gauge the propagator of the vector meson includes
three space-like polarization states. We notice that the most divergent
terms proportional to $m_{\rho }^{-2}$, which come from the longitudinal
component of the $\rho $ propagator, cancel. The term \ $\sim (k^{\mu
}q^{\nu }-g^{\mu \nu }q\cdot k)/D_{k}$ that remains in Eq.(\ref{Eq.B4}) is
equal to zero after integration over $k$. Further, it is straightforward to
check that the sum of the four diagrams does not depend on the photon gauge,
\textit{i.e.} \ $q_{\mu }(M_{1a}^{\mu \nu }+M_{1b}^{\mu \nu }+M_{1c}^{\mu
\nu }+M_{1d}^{\mu \nu })=0.$

To evaluate the integrals we apply the method of dimensional
regularization (see, e.g. \cite{Pes95}, App.A). Using the Feynman
parametrization and integrating over $k$ in $d$-dimensional space-time we obtain
\begin{eqnarray}
M_{1}^{\mu \nu } &=&2i f g^{\mu \nu } \frac{\Gamma (2-d/2)}{(4\pi )^{d/2}}
\left\{ - \frac{1}{2} \int_{0}^{1} \mbox{\rm d}u_{1} \int_{0}^{1-u_{1}}
\mbox{\rm d}u_{2} \, \Delta_{1}^{d/2-2}\right.
 \nonumber \\
&&\left. +\int_{0}^{1}\mbox{\rm d}u[\Delta _{2}^{d/2-2}-(1-\frac{u}{2}%
)\,\Delta _{3}^{d/2-2}]\right\} \, ,  \label{Eq.B6}
\end{eqnarray}
where the constant $f$ reads $f= -ie g_{\rho}^3 v\sqrt{X_{\pi}}(1+X_{\pi})/2$.
The expressions for $\Delta _{1},\Delta _{2},\Delta _{3}$ will be specified
below. We kept in Eq.(\ref{Eq.B6}) only the $g^{\mu \nu }$ part because of
the condition $Q\cdot \epsilon (\gamma )=0$ (in the rest frame of $a_{1}$).
Since the Gamma-function $\Gamma (2-d/2)$ has a pole at $d=4$ we have to
verify that the expression in the curly brackets vanishes at $d=4.$ Using
the expansions near $d=4$ \cite{Pes95}: $\ \Delta ^{d/2-2}=1+(d/2-2)\log
\Delta +\mathcal{O}((d/2-2)^{2})$ and \ $\Gamma (2-d/2)=2/(4-d)-\gamma _{E}+%
\mathcal{O}(d/2-2)$, \ we find that the pole term drops out because of the
relation
\begin{equation}
-\frac{1}{2}\int_{0}^{1}\mbox{\rm d}u_{1}\int_{0}^{1-u_{1}}\mbox{\rm d}%
u_{2}\,+\int_{0}^{1}\mbox{\rm d}u[1-(1-\frac{u}{2})]=0\,.
\label{Eq.B7}
\end{equation}
The residue gives the amplitude
\begin{equation}
M_{1}^{\mu \nu }=ifg^{\mu \nu }(4\pi )^{-2}\left\{
\int_{0}^{1}\int_{0}^{1}\log \,\Delta _{1}(x,y)\,\mbox{\rm d}x\mbox{\rm d}%
y+\int_{0}^{1}[(2-x)\,\log \Delta _{3}(x)-2\log \,\Delta _{2}(x)]\mbox{\rm d}%
x\right\} ,  \label{Eq.B8}
\end{equation}
where the integration variables $u_{1},\,u_{2}$ have been changed to $x,y$
defined such that $u_{1}=xy,\,\,u_{2}=x(1-y).$ In the above formulas $\gamma
_{E}\approx 0.5772$ is the Euler-Mascheroni constant and
\begin{eqnarray}
\Delta _{1}(x,y) &=&\Delta _{3}(x)-(m_{a}^{2}-m_{\pi
}^{2})\,x(1-x)y,\;\;\;\;\;\;\;\;\Delta _{3}(x)=m_{\rho }^{2}(1-x)+m_{\pi
}^{2}x^{2},\,  \nonumber \\
\Delta _{2}(x) &=&-m_{a}^{2}x(1-x)+m_{\rho }^{2}(1-x)+m_{\pi }^{2}x.
\label{Eq.B9}
\end{eqnarray}
The argument of the logarithm can formally be made dimensionless by changing
$\ \log \Delta _{i}\rightarrow \log (\Delta _{i}/\Lambda ^{2}),$
where $\Lambda $ is a mass-scale parameter.
This will not affect the amplitude, due to (\ref{Eq.B7}).
Expression (\ref{Eq.B8}) can be further
simplified by carrying out the integration over $y.$

The $N\bar{N}$-loop amplitudes [$j=8$ in Fig.~\ref{fig:2}] require
a trace calculation, for example,
\begin{eqnarray}
M_{8a}^{\mu \nu } &\Longrightarrow &\frac{\mbox{Tr}
[(k\hspace{-0.5em}/+p\hspace{-0.5em}/+m_{N})\gamma ^{\mu }
(k\hspace{-0.5em}/+Q\hspace{-0.5em}/+m_{N})\gamma ^{\nu }\gamma _{5}
(k\hspace{-0.5em}/+m_{N})\gamma _{5}]}{D_{k}\ D_{k+Q}D_{k+p}}
 \nonumber \\
&\Longrightarrow &4m_{N}\left[ \frac{g^{\mu \nu }q\cdot (2k+Q)-(2k+Q)^{\mu
}q^{\nu }}{D_{k}\ D_{k+Q}D_{k+p}}-\frac{g^{\mu \nu }}{D_{k}\ D_{k+p}}\right],
 \label{Eq.B10}
\end{eqnarray}
where $D_{k}=k^{2}-m_{N}^{2},\,\ D_{k+Q}=(k+Q)^{2}-m_{N}^{2},\,\ $and$\ \
D_{k+p}=(k+p)^{2}-m_{N}^{2}$. The $(c)$ amplitude in Fig.~\ref{fig:2}
reads
\begin{equation}
M_{8c}^{\mu \nu }\Longrightarrow 4m_{N}\frac{g^{\mu \nu }}{D_{k}\ D_{k+p}} .
\label{Eq.B11}
\end{equation}
This amplitude cancels the divergent and EM gauge noninvariant piece
of $M_{8a}^{\mu \nu }.$ \ Finally, the amplitude $M_{8d}^{\mu \nu }$ (radiation
from the pion line) is zero, being proportional to $Q^{\nu }.$

An important observation is the cancellation of divergent terms between
different diagrams. This, together with the EM gauge invariance, helps in
evaluating the other diagrams in Fig.~\ref{fig:2}. Most of these diagrams are
calculated similarly to $M_{1}^{\mu \nu }$ while others, where the photon couples to
a vector meson in the loop, are more algebraically involved.

After this general consideration we present expressions for the amplitudes
corresponding to the diagrams in Fig.~\ref{fig:2} with $j=1,...,8$.
For the $\pi -\rho $ loops ($j=1,2$) we obtain
\begin{eqnarray}
T_{1} &=&C_{1}\{-\frac{1}{2}-\int_{0}^{1}\log \frac{\Delta _{2}(x)}{\Delta
_{3}(x)}[2-x+\frac{\Delta _{3}(x)}{M^{2}(1-x)}]\ \mbox{\rm d}x\},
\label{Eq.B12} \\
T_{2} &=&C_{2}\{\frac{1}{2}+\int_{0}^{1}\log \frac{\Delta _{2}(x)}{\Delta
_{3}(x)}[x-3+\frac{1}{x}(\frac{\Delta _{3}(x)}{M^{2}}+2)]\ \mbox{\rm d}x\},
\label{Eq.B13} \\
\Delta _{2}(x) &=&m_{\pi }^{2}x+m_{\rho }^{2}(1-x)-m_{a}^{2}x(1-x),\,\ \ \ \
\ \ \ \ \ \ \Delta _{3}(x)=m_{\pi }^{2}x^{2}+m_{\rho }^{2}(1-x),  \nonumber
\\
C_{1} &=&C_{2}=(4\pi )^{-2}\frac{1}{2}eg_{\rho }^{3}v\sqrt{X_{\pi }}
(1+X_{\pi }),  \nonumber
\end{eqnarray}
and we introduced the notation $M^{2}\equiv m_{a}^{2}-m_{\pi }^{2}=2Q\cdot q.
$ \ Since \ $m_{\pi }+m_{\rho }<m_{a}$ \ these amplitudes acquire an imaginary
part if $\ \Delta _{2}(x)<0,$ and we have to select the proper branch of the
logarithm. Recalling the prescription $\ m^{2}\rightarrow m^{2}-i0$ \ for
the masses of the particles in the propagators, we have correspondingly $\
\Delta _{2}(x)\rightarrow \Delta _{2}(x)-i0$ and we can use the substitution
\begin{eqnarray}
\log \Delta _{2}(x) &=&\log |\Delta _{2}(x)|-i\pi \theta (-\Delta
_{2}(x))=\log |\Delta _{2}(x)|-i\pi \theta (x-x_{-})\theta (x_{+}-x),
\label{Eq.B14} \\
x_{\pm } &=&\{(m_{a}^{2}+m_{\rho }^{2}-m_{\pi }^{2})\,\pm \lbrack
(m_{a}^{2}+m_{\rho }^{2}-m_{\pi }^{2})^{2}-4m_{a}^{2}m_{\rho
}^{2}]^{1/2}\}/2m_{a}^{2} , \nonumber
\end{eqnarray}
in Eq.(\ref{Eq.B12}) and (\ref{Eq.B13}) to calculate the real and imaginary
parts. Here $\theta (y)=1$ if$\ y>0,$ and $0$ otherwise.

For the $j=3$ contribution we obtain
\begin{eqnarray}
T_{3} &=&C_{3}\{-1+2\int_{0}^{1}\log \frac{\Delta _{2}(x)}{\Delta _{3}(x)}[x-%
\frac{\Delta _{3}(x)}{M^{2}(1-x)}]\ \mbox{\rm d}x\},
\label{Eq.B15}\\
\Delta _{2}(x) &=&m_{\pi }^{2}x+m_{\sigma }^{2}(1-x)-m_{a}^{2}x(1-x),\,\ \ \
\ \ \ \ \ \ \ \ \ \ \ \Delta _{3}(x)=m_{\pi }^{2}x^{2}+m_{\sigma }^{2}(1-x),
\nonumber \\
C_{3} &=&(4\pi )^{-2}2eg_{\rho }\lambda v\sqrt{X_{\pi }}X_{\pi },\,\ \ \ \ \
\ \ \ \ \ \ \ \ \ \ \ \ \ \ \ \ \ \ \ \ \ \ \ \ 2\lambda =(m_{\sigma
}^{2}-m_{\pi }^{2})/v^{2}\,.  \nonumber
\end{eqnarray}
A substitution similar to Eq.(\ref{Eq.B14}) is applied to calculate the real and
imaginary part of $T_{3}.$

Next, for the diagrams with $j=4$ in Fig.~\ref{fig:2} the amplitude is real and reads,
\begin{eqnarray}
T_{4} &=&C_{4}\{\frac{1}{2}(1-\delta )+\int_{0}^{1}\log \frac{\Delta _{2}(x)
}{\Delta _{3}(x)}[x-3+\delta (1-x)+\frac{1}{x}(\frac{\Delta _{3}(x)}{M^{2}}
(1-\delta )+2)]\ \mbox{\rm d}x\},
\label{Eq.B16}\\
\Delta _{2}(x) &=&m_{\sigma }^{2}x+m_{a}^{2}(1-x)^{2},\,\ \ \ \ \ \ \ \ \ \
\ \ \ \ \ \ \ \Delta _{3}(x)=-m_{\pi }^{2}x(1-x)+m_{\sigma
}^{2}x+m_{a}^{2}(1-x),  \nonumber \\
\ C_{4} &=&(4\pi )^{-2}2eg_{\rho }^{3}v\sqrt{X_{\pi }},\,\ \ \ \ \ \ \ \ \ \
\ \ \ \ \ \ \ \ \ \ \ \delta =(m_{\sigma }^{2}-m_{\pi }^{2})/m_{a}^{2}\ .
\nonumber
\end{eqnarray}

The contribution $j=5$ in Fig.~\ref{fig:2} with
an intermediate Higgs meson $\eta $ ($\pi ^{+} \eta$ state) is
\begin{eqnarray}
T_{5} &=&C_{5}\{-1+2\int_{0}^{1}\log \frac{\Delta _{2}(x)}{\Delta _{3}(x)}
[x-\frac{\Delta _{3}(x)}{M^{2}(1-x)}]\ \mbox{\rm d}x\},\,
\label{Eq.B17} \\
\Delta _{2}(x) &=&m_{\pi }^{2}x+m_{H}^{2}(1-x)-m_{a}^{2}x(1-x),\,\ \ \ \ \ \
\ \ \ \ \ \ \Delta _{3}(x)=m_{\pi }^{2}x^{2}+m_{H}^{2}(1-x),  \nonumber \\
C_{5} &=&-(4\pi )^{-2}\frac{1}{4}e\lambda _{H}m_{\rho }\sqrt{1-X_{\pi }}
(1-X_{\pi }),\,  \nonumber
\end{eqnarray}

The amplitudes for $j=6$ ($a^{+} \eta$ state) and $j=7$ ($\rho ^{+} \zeta$
state) can be written in the form
\begin{equation}
T_{6/7}=C_{6/7}\{\frac{1}{2}(1-\delta )+\int_{0}^{1}\log \frac{\Delta _{2}(x)%
}{\Delta _{3}(x)}[x-3+\delta (1-x)+\frac{1}{x}(\frac{\Delta _{3}(x)}{M^{2}}%
(1-\delta )+2)]\ \mbox{\rm d}x\},
\label{Eq.B18}
\end{equation}
where
\begin{eqnarray}
\Delta _{2}(x) &=&m_{H}^{2}x+m_{a}^{2}(1-x)^{2},\,\ \ \ \ \ \ \ \ \ \ \ \ \
\ \ \ \ \Delta _{3}(x)=-m_{\pi }^{2}x(1-x)+m_{H}^{2}x+m_{a}^{2}(1-x),
\nonumber \\
C_{6} &=&-(4\pi )^{-2}\frac{1}{4}eg_{\rho }^{3}u\sqrt{1-X_{\pi }},\,\ \ \ \
\ \ \ \ \ \ \ \ \ \ \ \ \delta =(m_{H}^{2}-m_{\pi }^{2})/m_{a}^{2}
\nonumber
\end{eqnarray}
for $j=6,$ and
\begin{eqnarray}
\Delta _{2}(x) &=&m_{H}^{2}x+m_{\rho }^{2}(1-x)-m_{a}^{2}x(1-x),\,\ \ \ \ \
\ \Delta _{3}(x)=m_{H}^{2}x+m_{\rho }^{2}(1-x)-m_{\pi }^{2}x(1-x),  \nonumber
\\
C_{7} &=&C_{6},\,\;\;\;\;\;\;\;\;\;\;\;  \ \ \ \ \delta
=(m_{H}^{2}-m_{\pi }^{2})/m_{\rho }^{2}  \nonumber
\end{eqnarray}
for $j=7.$ In the limit of large Higgs mass, $m_{H}^{2}\gg m_{\pi
}^{2},m_{\rho }^{2},m_{a}^{2},$ \ the sum of the amplitudes with the
intermediate $\eta $ and $\zeta $ mesons can be written as
\begin{equation}
T_{5}+T_{6}+T_{7}\approx (4\pi )^{-2}eg_{\rho }^{2}\frac{m_{a}^{2}-m_{\pi
}^{2}}{6 m_{\rho }} \sqrt{1-X_{\pi }}+\mathcal{O}(m_{H}^{-2}).
\label{Eq.B19}
\end{equation}
This amplitude corresponds to a contact-like Lagrangian in Eq.(\ref{Eq.54})
of sect.~\ref{subsubsec:a-pi-gam}.

Finally, the amplitude for
$j=8$ ($p\bar{n}$ state) on Fig.~\ref{fig:2} has the form
\begin{eqnarray}
T_{8} &=&C_{8}\int_{0}^{1}\log \left[ \frac{m_{N}^{2}-m_{a}^{2}x(1-x)}{%
m_{N}^{2}-m_{\pi }^{2}x(1-x)}\right] \frac{(1-2x)\mbox{\rm d}x}{1-x}\ ,
\label{Eq.B20} \\
C_{8} &=&(4\pi )^{-2}2eg_{\rho }g_{\pi }m_{N}\sqrt{X_{\pi }}\ .
\nonumber
\end{eqnarray}

The amplitude on Fig.~\ref{fig:3} describing the decay $\rho ^{+}\rightarrow
\pi ^{+} \gamma $ is calculated similarly. Evaluating the trace we obtain
\begin{eqnarray}
M^{\mu \nu } &\Longrightarrow &\frac{\mbox{Tr} [(k \hspace{-0.5 em} /+p
\hspace{-0.5 em} /+m_{N})\gamma ^{\mu }(k \hspace{-0.5 em} /+Q \hspace{-0.5
em} /+m_{N})\gamma ^{\nu }(k \hspace{-0.5 em} /+m_{N}) \gamma _{5}]}{D_{k}\
D_{k+Q}D_{k+p}}  \nonumber \\
&\Longrightarrow &4m_{N}\,\varepsilon ^{\mu \nu \lambda \sigma }Q_{\lambda
}q_{\sigma }\frac{1}{D_{k}\ D_{k+Q}D_{k+p}}.
\label{Eq.B21}
\end{eqnarray}
The result of the integration over $k$ is given
in Eq.(\ref{Eq.70}) of Sect.~\ref{subsubsec:rho-pi-gam}.

%%%%%%%%%%%%%%%%%%%%%%%%%%% REFERENCES %%%%%%%%%%%%%%%%%%%%%%%%%%

\end{document}